\documentclass[aps,prd,twocolumn,showpacs,groupedaddress,nofootinbib]{revtex4}
\usepackage{graphicx}  
\usepackage{dcolumn}   
\usepackage{bm}        
\usepackage[english]{babel}
\usepackage{amsfonts,amsmath,amssymb,mathrsfs}
\usepackage{epstopdf}
\usepackage{subfigure}
\usepackage{enumerate}

\usepackage{latexsym}
\def\beq{\begin{equation}}
\def\eeq{\end{equation}}
\def\bea{\begin{eqnarray}}
\def\eea{\end{eqnarray}}
\def\ben{\begin{enumerate}}
\def\een{\end{enumerate}}

\def\a{\varphi}
\def\r{\rho}
\def\s{\sigma}
\def\t{\tau}

\def\th{\theta}
\def\G{\Gamma}

\def\l{\lambda}
\def\D{\Delta}

\def\o{\omega}\def\O{\Omega}
\def\d{\delta}

\def\tS{\tilde{\Sigma}}

\def\td{\dot{t}}
\def\rd{\dot{r}}

\def\thd{\dot{\theta}}
\def\Xd{\dot{X}}

\def\half{\textstyle{\frac{1}{2}}}
\begin{document}

\title{String dynamics and ejection along the axis of a spinning black hole}
\author{Ted Jacobson and Thomas P. Sotiriou}
\affiliation{Center for Fundamental Physics,  University of Maryland, College Park, MD 20742-4111, USA}

\begin{abstract} 

Relativistic current carrying strings moving axisymmetrically
on the background of a Kerr black hole are studied. The 
boundaries and possible types of motion of a string with 
a given energy and current are found.
Regions of parameters for which the string falls into the black hole, or is trapped 
in a toroidal volume,  or can escape to infinity, are identified, and 
representative trajectories are examined by numerical integration, illustrating 
various interesting behaviors. In particular, we find that a string can start out at
rest near the equatorial plane and, after bouncing around, be ejected out along the axis,
some of its internal (elastic or rotational kinetic) energy having been transformed into 
translational kinetic energy. The resulting velocity can be an order unity fraction of the
speed of light. This process results from the presence of an outer tension barrier
and an inner angular momentum barrier that are deformed by the gravitational field. 
We speculatively discuss the possible astrophysical significance of this mechanism
as a means of launching a collimated jet of MHD plasma flux tubes along the 
spin axis of a gravitating system fed by an accretion disk. 

\end{abstract} 

\maketitle

\section{Introduction}

In this paper we study the motion of an axisymmetric, current carrying relativistic string
loop in the background of a Kerr black hole. This interesting dynamical system
has been studied before in its own right~\cite{Larsen:1993nt,Frolov:1999pj}), 
and our goal here is to better understand its behavior.

Our investigation is loosely motivated
by the problem of the production of collimated 
astrophysical jets 
of matter. Systems that exhibit collimated jets range from accreting young stars, neutron stars and black holes to supermassive black holes 
in quasars and
active galactic nuclei~\cite{livio}. 
Magnetized plasmas interacting with an accretion disk are believed to play 
a central role in
jet production, but despite much work and many prosposals for the mechanism,
the process is still not yet well understood.
Such plasmas are governed by magnetohydrodynamics
(MHD), a complicated nonlinear field theory. Under some circumstances, plasmas
exhibit associated string-like behavior, either via the dynamics of magnetic 
field lines embedded in the plasma~\cite{Christensson:1999tp,semenov,Semenov:2004ib}, or in the dynamics of relatively thin isolated 
flux tubes of plasma which can be approximately described by an effective  
one-dimensional string~\cite{spruit}. 
In either of these cases, the string would be described by an 
energy density
and a tension, and might be carrying currents of mass and/or charge. 
The hope is that some essential aspects of the physics could be
captured by string dynamics, which is tremendously simpler than MHD.

The dynamical
formalism for describing relativistic current carrying strings is
well-developed~\cite{Hindmarsh:1994re,vilshell,Larsen:1993ha,Carter:1996rn},
and a generalization to a wider variety of equations of state 
is given in Ref.~\cite{Carter:1997pb}.
Here we restrict attention to a relatively simple class of strings, namely 
axisymmetric loops characterized by a constant tension and a conserved 
mass current on the string worldsheet,
in order to begin developing some insight into the factors influencing the string dynamics.
This could be generalized in future work to allow for different equations of state, 
coupling of a charge current to an external 
electromagnetic field~\cite{Larsen:1992xz,Larsen:1992ya}, and deviations from 
axisymmetry~\cite{Dyadechkin:2008zz, Dubath:2006vs, Snajdr:2002aa, Snajdr:2002rw}. 

The string tension prevents a string loop from expanding beyond some radius,
and the worldsheet current can produce an angular momentum barrier preventing 
the  loop from collapsing into the black hole. Thus, depending on its energy and current, 
a string may be trapped in a toroidal region surrounding the black hole, or its motion may be 
confined to a cylindrical shell that extends to infinity. In the latter case, one
can envisage processes wherein internal energy of the string is converted into 
translational kinetic energy via the propagation on the black hole background. 
This is one of the
phenomena we aim to understand. The other key question is what role may be 
played by the spin of the black hole, as a result of the associated dragging
of inertial frames.

\section{Current carrying string in curved spacetime}
\label{covfeq}

The string worldsheet is described by giving its spacetime coordinates
$X^\l(\s^a)$ ($\l=0,1,2,3$)  
as functions of the two worldsheet coordinates $\s^a$ ($a=0,1$).
The induced metric on the worldsheet is
\beq
\label{induced}
h_{ab}= g_{\mu\nu}X^\mu_{,a}X^\nu_{,b},
\eeq
where $g_{\mu\nu}$ is the ambient spacetime metric,
for which we adopt the signature ($-$$+$$+$$+$).
To describe the current we introduce a scalar field $\a(\s^a)$ 
living on the worldsheet. 
We consider dynamics generated by the action
\begin{equation}
\label{scs}
S=-\int d^2\s\, \sqrt{-h}\left(\mu/c+h^{ab}\a_{,a} \a_{,b}\right),
\end{equation}
where the constant $\mu$ denotes the string tension, 
and $c$ is the speed of light which is
hereafter set to unity.
We assign the line element $g_{\mu\nu} dx^\mu dx^\nu$ 
the dimensions of length squared, and take the 
worldsheet coordinates to be dimensionless, 
so that the dimensions of $h_{ab}$ are length squared,
those of $\mu$ are energy over length,
and those of $\varphi$ are square root of action.\footnote{Strings described 
by action (\ref{scs}) were originally introduced~\cite{Witten:1984eb} 
as an effective description of 
``superconducting strings", a type of topological defect that
might occur in a theory with multiple scalar fields undergoing spontaneous
symmetry breaking.}

The action is the integral of a worldsheet scalar density,
hence it is independent of the choice of worldsheet 
coordinates. Note that the part of the action involving the scalar field $\a$ is 
invariant under conformal rescalings of the metric. 
One could of course consider
other scalar functions of the invariant $h^{ab}\a_{,a} \a_{,b}$ in
the Lagrangian, but for the exploratory purposes of the present
investigation the simple choice in (\ref{scs}) is sufficient.
In any case, one ultimately wants to generalize the system to
allow for a wider class of equations of state that are perhaps more
suitable for MHD applications. 

Varying the action with respect to $h_{ab}$ yields the worldsheet
stress-energy tensor density $\tS^{ab}$,
\beq
\d_h S = \half\int d^2\s\, \tS^{ab}\d h_{ab},
\eeq
where
\beq\label{tS}
\tS^{ab}=\sqrt{-h}\Bigl(2j^a j^b - (\mu+j^2)h^{ab}\Bigr),
\eeq
and 
\beq
j_a = \a_{,a}, \qquad j^a = h^{ab}j_b, \qquad j^2 = h^{ab}j_aj_b.
\eeq
The tilde on $\tS^{ab}$ is a reminder that this quantity has density weight one
with respect to worldsheet coordinate transformations. 
Because of the conformal invariance, the $j$-dependent part of $\tS^{ab}$ is traceless.
The contribution from the string tension
is proportional to the metric, and with $\mu>0$ has 
a positive energy density and
an opposite, negative pressure, i.e.\ a tension. 
The contribution from the 
current is traceless, due to the 
conformal invariance of that part of the action. 
It can be viewed as a 1+1 dimensional massless radiation fluid,
with positive energy density and equal pressure.

Varying the action with respect to $X^\mu$ yields the equation of 
motion\footnote{One could rewrite eq.~(\ref{eom1}) in terms of the 
Christoffel symbols of $g_{\mu\nu}$, however
it is more convenient to work with the explicit form (\ref{eom1}).}
\beq
\label{eom1}
(\tS^{ab}g_{\mu\l}X^\mu_{,a})_{,b}-\half \tS^{ab}g_{\mu\nu,\l} X^\mu_{,a} X^\nu_{,b}=0.
\eeq
%
%
Varying the action with respect to $\a$ yields the 1+1 dimensional
wave equation,
\beq
\label{eomj}
(\sqrt{-h} h^{ab}\a_{,a})_{,b}=0,
\eeq
which also implies that the current is divergenceless, $(\sqrt{-h}j^a)_{,a}=0$.
The worldsheet stress-energy tensor is also divergenceless, 
$\tS^{ab}{}_{;b}=0$, with respect to the $h_{ab}$-compatible
covariant derivative. (To verify this directly  one must use $j_{a,b}=j_{b,a}$,
which follows from $j_a=\a_{,a}$. Conversely, $(\sqrt{-h}j^a)_{,a}=0$
and $\tS^{ab}{}_{;b}=0$, together imply that $j_a=\a_{,a}$ for some scalar field 
$\a$.)        

\section{Axisymmetric dynamics in a stationary, axisymmetric spacetime}
\label{dynamics}

\subsection{Conformal gauge}
Every two dimensional metric is conformal to a locally flat metric,
hence in particular
\beq\label{O}
h_{ab}=\O^2 \eta_{ab},
\eeq
where $\eta_{ab}$ is locally flat and
$\O$ is a worldsheet scalar function.
We can always adopt coordinates 
$\s^a=(\t,\s)$
such that the coordinate components
of $\eta_{ab}$ are $\eta_{\t\s}=0$ and $\eta_{\t\t}=-\eta_{\s\s}=-1$, 
i.e.\ the constant Minkowski metric.
This ``conformal gauge" choice
is equivalent to requiring $h_{\t\s}=0$ and $h_{\t\t}+h_{\s\s}=0$,
which as conditions on $X^\mu$ read
\beq\label{confgauge}
g_{\mu\nu}X^\mu_{,\t}X^\nu_{,\s}=0
=g_{\mu\nu}(X^\mu_{,\t}X^\nu_{,\t}+X^\mu_{,\s}X^\nu_{,\s}).
\eeq
In this gauge 
$h=-\O^4$, and the metric components
satisfy $\sqrt{-h}\, h^{ab}=\eta^{ab}$,
where $\eta^{ab}$ is the inverse of the Minkowski
metric. The conformal gauge conditions (\ref{confgauge})
do not completely fix the coordinates: the lightlike
combinations $\t\pm\s$ may be replaced by any smooth
monotonic functions of themselves.

We aim to study axisymmetric string motion in  
an axisymmetric, stationary spacetime. Such a
spacetime is 
described by a metric, written in coordinates $(t,r,\theta,\phi)$,
of the general form
\beq \label{ds2}
ds^2 = g_{tt}dt^2 + 2g_{t\phi}dt d\phi
+g_{\phi\phi}d\phi^2 +g_{rr}dr^2
+g_{\theta\theta}d\theta^2 
\eeq
where all the metric components are independent of $t$ and $\phi$.

It is tempting to choose the worldsheet coordinate $\s$ to be 
equal to the spacetime coordinate $\phi$, in which case
a general axisymmetric worldsheet would take the form
$X^\mu(\s,\t)=(t(\t),r(\t),\theta(\t),\s)$. However, this 
would imply $h_{\t\s}=g_{t\phi}t_{,\t}$, which  
does not satisfy the conformal gauge condition 
if $g_{t\phi}\ne0$. We therefore allow for a $\t$ dependent relative shift 
$\phi=\s+f(\t)$, so that
\beq\label{X}
X^\mu(\s,\t)=(t(\t),r(\t),\theta(\t),\s+f(\t)).
\eeq
The gauge condition $h_{\t\s}=0$ then becomes
\beq\label{Xd-phi}
g_{\mu\phi}\Xd^\mu=g_{t\phi} \td + g_{\phi\phi} \dot{f}=0,
\eeq
which determines $f$ via
\beq
 \dot{f}=-(g_{t\phi}/g_{tt}) \td,\label{fdot}
\eeq
where the dot stands for $d/d\t$.
Equation (\ref{Xd-phi}) means that 
$\Xd^\mu$, and thus the worldsheet vector 
$\partial_\t$, are orthogonal to the 
axial rotation Killing vector $\partial_\phi$.
That is, the curves of constant $\s$ are
zero angular momentum (ZAMO) worldlines. 
For later reference we note that this
same gauge condition yields the useful relation
\beq\label{Xd-t}
g_{\mu t}\Xd^\mu=(g_{tt}-g_{t\phi}^2/g_{\phi\phi})\td. 
\eeq

The second gauge condition $h_{\t\t}+h_{\s\s}=0$
becomes
\beq
(g_{tt}-g_{t\phi}^2/g_{\phi\phi})\td^2+g_{rr}\rd^2+g_{\theta\theta}\thd^2+g_{\phi\phi}=0.
\label{gauge2}
\eeq
The conformal factor $\O^2$ is determined in this gauge 
by the equation $h_{\s\s}=\O^2\eta_{\s\s}=\O^2$, i.e.
\beq\label{Omega}
\O^2=g_{\phi\phi}.
\eeq

\subsection{Equations of motion}

In the conformal gauge (\ref{O}) 
the equation of motion (\ref{eomj}) for $\a$ takes the simple form
\beq\label{waveeqn}
\a_{,\t\t}-\a_{,\s\s}=0.
\eeq
The general solution is
$\a=\a_+(\t+\s)+\a_-(\t-\s)$, where $\a_\pm$ are arbitrary functions. 
On the other hand, 
the assumption of 
axisymmetry implies that the current is
independent of $\s$. Thus $j_{a,\s}=0$, which implies
$ \a_{,\t\s}=\a_{,\s\s}=0$. Together with (\ref{waveeqn}) this
implies that $\a$ must be linear in both $\s$ and $\t$,
\beq\label{alpha}
\a=j_\s\s+ j_\t\t,
\eeq
where $j_\s$ and $j_\t$ are constants.\footnote{Note that while
$j_s$ is single-valued, the scalar $\a$ is multi-valued. When the 
string action is taken to be an effective description of a field theory
defect~\cite{Witten:1984eb}, $\a$ is a phase defined modulo 
$2\pi$, which implies that the possible values of the current are
quantized.} 
Then $j^2=\Omega^{-1}(j_\s^2-j_\t^2)$, 
and the components of $\tS^{ab}$ (\ref{tS}) become
\bea\label{Sigmatilde}
\tS^{\t\t}&=& g_{\phi\phi}^{-1}(j_\s^2+ j_\t^2)+\mu\label{tStt}\\
\tS^{\s\s}&=&g_{\phi\phi}^{-1}(j_\s^2+ j_\t^2)-\mu\label{tSss}\\
\tS^{\s\t}&=&-2g_{\phi\phi}^{-1}j_\s j_\t.\label{tSst}
\eea
%

The quantity $\tS^{\t\t}$ in the conformal gauge 
is equal to the string energy density 
as measured by observers co-moving with the string at
constant $\s$. 
These are the co-moving ZAMO observers
mentioned in the discussion following eq.~(\ref{fdot}).
To verify 
the claim about the relation between $\tS^{\t\t}$ and the string energy density, note that 
since the 2-velocity $u^a$ of those observers is 
orthogonal to the constant $\tau$ surfaces, we have
$u_a=\O \partial_a \tau$, the prefactor $\O$ being
determined by the unit normalization of $u^a$.
Also $h=-\O^4$. The energy density is thus 
given by $(-h)^{-1/2}\tS^{ab}u_a u_b=\tS^{\t\t}$.
In particular, $\mu$ is the contribution form the 
string tension, and $(j_\s^2+ j_\t^2)/g_{\phi\phi}$ is the
energy density of the current.

The equations of motion (\ref{eom1}) can now be written out 
more explicitly in our chosen gauge. Using the gauge conditions
and the axisymmetry, they take the form
\begin{align}
\label{eom1a}
&(\tS^{\t\t}g_{\mu\l}\Xd^\mu + \tS^{\s\t}g_{\phi\l})_{,\t}\\
&\quad-\half(\tS^{\t\t}g_{\mu\nu,\l}\Xd^\mu\Xd^\nu+\tS^{\s\s}g_{\phi\phi,\l}+\tS^{\t\s}g_{t\phi,\l}\,\dot{t})=0.\nonumber
\end{align}
For $\l$ equal to $\phi$ 
this equation is identically satisfied. For $\lambda=t$,  it yields 
(using (\ref{Xd-t})) the energy conservation law,
\beq\label{eomt}
\tS^{\t\t}(g_{tt}-g_{t\phi}^2/g_{\phi\phi})\td + \tS^{\s\t}g_{t\phi}=-E,
\eeq
where $E$ is a constant to be identified below with the total Killing energy 
of the string (divided by $2\pi$).
For $\l$ equal to $r$ or $\theta$ it yields
\begin{align}
 \left(\tS^{\t\t}g_{rr}\rd\right)_{,\t} -&\Big(\tS^{\t\t}g_{\mu\nu,r}\Xd^\mu\Xd^\nu+\tS^{\s\s}g_{\phi\phi,r}\Big)=0\label{eomr}\\
\left(\tS^{\t\t}g_{\th\th}\thd\right)_{,\t} -&\Big(\tS^{\t\t}g_{\mu\nu,\th}\Xd^\mu\Xd^\nu +\tS^{\s\s}g_{\phi\phi,\th}\Big)=0.\label{eomth}
\end{align}

The dynamics depends on the current only through the worldsheet
stress tensor (\ref{tStt}-\ref{tSst}), in which the current enters only
in the 
two quadratic combinations $j_\s^2+ j_\t^2$ and $j_\s j_\t$,
both of which are symmetric under interchange of 
$j_\s$ and $j_\t$.
Note that $j_\s j_\t $ appears only via $\dot{t}$, through the 
energy conservation law 
(\ref{eomt}), and then only when $g_{t\phi}\ne0$.

To parametrize the solutions, we will later use the first of
these combinations, together with the ratio of the current components,
\bea
J^2&\equiv &j_\s^2+ j_\t^2\\
\o &\equiv &-j_\s/j_\t.
\eea
(The minus
sign is included so that positive $\o$ will
correspond to positive angular momentum.
Note that $-j_\s/j_\t=j^\s/j^\t$.)
The product $j_\s j_\t $ can be expressed
in terms of $J$ and $\o$ 
as $j_\s j_\t = -\o J^2/(1+\o^2)$.
Since the dynamics is 
symmetric under interchange of $j_\s$ and $j_\t$, 
the range $-1\le\o\le1$ covers all distinct 
cases. The extreme values $\o=\pm1$ corresond
to null currents. 

\subsection{Conserved quantities}

Suppose $\xi^\l$ is a spacetime Killing vector. If the
spacetime coordinates are chosen so the components of $\xi^\l$
are constant everywhere, then Killing's
equation implies $g_{\mu\nu,\lambda} \xi^\lambda = 0$.
Contracting (\ref{eom1}) with $\xi^\l$ then
yields a conserved Killing current. That is,
the worldsheet vector density
\beq
\label{killcur}
{\cal J}_\xi^b = \tS^{ab} X^{\mu}_{,a} g_{\mu\l} \xi^\l
\eeq
satisfies
\beq
\label{killcur2}
{\cal J}_\xi^b{}_{,b}=0.
\eeq
Although we argued for its conservation using an adapted coordinate system,
${\cal J}_\xi^b$ is manifestly spacetime coordinate independent.  
Note that it is just the worldsheet energy-momentum tensor contracted
with the pullback of the Killing one-form to the worldsheet. Thus
integrating it over any closed cross section of the worldsheet gives
the conserved quantity associated with the Killing vector, 
\beq
 Q_\xi=\int {\cal J}_\xi^b dS_b.
\eeq
If we take the integral over a surface of constant $\t$  
and use the axisymmetry, the integral becomes simply
\beq
Q_\xi=2\pi{\cal J}_\xi^\t=2\pi \tS^{a\tau}X^\mu_{,a} g_{\mu\l}\xi^\l.
\eeq

When $\xi^\l$ is the Killing vector $\partial_t$ corresponding to $t$ translation symmetry, then
(\ref{X}) and (\ref{Xd-t}) yield
\beq
-Q_{\partial_t}={\cal E}=2\pi E,
\eeq
where $E$ is the constant introduced in eq.~(\ref{eomt}). 
Thus $E$ is in
fact the Killing energy of the string divided by $2\pi$ .
When $\xi^\l$ is the Killing vector $\partial_\phi$ 
corresponding to $\phi$ translation symmetry (rotation), then
using (\ref{X}), (\ref{Xd-phi}) and (\ref{tSst}) 
we find the angular momentum $L$ of the string,
\beq
Q_{\partial_\phi}=L=-4\pi  \, j_\s j_\t.
\eeq
This is manifestly constant for the solutions (\ref{alpha}) under consideration.
Without the current, the string carries no angular momentum, since the
stress tensor is then proportional to $h_{ab}$, which is
boost invariant along the worldsheet. 

\subsection{Effective potential}

We can solve (\ref{eomt}) for $\td$ 
and substitute into the gauge condition (\ref{gauge2}), yielding
\beq\label{gaugeKerr}
(g_{t\phi}^2/g_{\phi\phi}-g_{tt})(\tS^{\t\t})^2(g_{rr}\rd^2+g_{\th\th}\thd^2) + V(r,\theta)=0,
\eeq
where 
\beq\label{VKerr}
V(r,\th)=-(E+g_{t\phi}\tS^{\s\t})^2+(g_{t\phi}^2-g_{tt}g_{\phi\phi})(\tS^{\t\t})^2
\eeq
is what we call here the ``effective potential".   
The motion outside the horizon 
is confined to the region where $V(r,\th)\le0$, since 
the first term in (\ref{gaugeKerr}) is non-negative 
outside the horizon.\footnote{The determinant
of the metric (\ref{ds2}) is 
$(g_{tt}g_{\phi\phi}-g_{t\phi}^2)g_{rr}g_{\th\th}$, which must
be negative for a Lorentzian metric. For the metrics we consider,
$g_{\th\th}>0$ everywhere and $g_{rr}, g_{\phi\phi}>0$
outside the horizon,  so evidently $g_{t\phi}^2/g_{\phi\phi}-g_{tt}$ is positive
outside the horizon.}
More explicitly, the motion is bounded by the contour\footnote{To obtain 
this we have taken a square root, but the root
is unique since, according to (\ref{eomt}) the combination
$E+g_{t\phi}\tS^{\s\t}$ is positive for trajectories outside
the horizon.} 
\bea\label{VKerr0}
E=E_b&\equiv&(g_{t\phi}^2-g_{tt}g_{\phi\phi})^{1/2}\, 
\tS^{\t\t}-g_{t\phi}\tS^{\s\t}\\
&=&A(r,\theta)\mu + B(r,\theta,\o) J^2, 
\eea
where $A$ and $B$ are functions determined by the metric components and, in the case of $B$, the value of $\o$.
The initial conditions and the value of the current
determine $E$ through (\ref{gauge2}) and (\ref{eomt}), and the
subsequent motion is confined to the region contained within
the contour (\ref{VKerr0}).  We shall determine that motion 
in several representative cases by numerical 
integration.

\section{Newtonian elastic ring}
\label{newtanal}

Before presenting the results for the relativistic string, we 
develop a Newtonian analogue of the system, to help interpret the
dynamics. Indeed, it turns out that most of the salient features
are independent of relativistic effects.

Consider a Newtonian elastic ring of total mass $m$, moving axisymmetrically in the
gravitational field of a point mass $M$. Using cylindrical coordinates $(\rho,z,\phi)$,
the conserved Newtonian energy of the system is given by 
\beq
E_N= \half m (\dot{\r}^2+\dot{z}^2 + \r^2 \o^2)+\half k \r^2-GMm/(\r^2+z^2)^{1/2}\, ,
\eeq
where $\o$ is the angular velocity and $k$ the elastic constant. 
The angular momentum $l=m\r^2\o$ about the axis is conserved.
Using $l$,
energy conservation can be expressed as
\beq
\half m (\dot{\r}^2+\dot{z}^2)+V_N(\r,z) =0,
\eeq
with
\beq\label{VN}
V_N(\r,z)=-E_N + \frac{l^2}{2m\r^2} + {\half k \r^2}-\frac{GMm}{\sqrt{\r^2+z^2}}.
\eeq
The motion of the ring is confined to the region
$V_N(\r,z)\le0$, and thus bounded by the contour 
$V_N(\r,z)=0$, i.e. the curve in the $\r$-$z$ plane given by 
\beq\label{Ncontour}
E_N = \frac{l^2}{2m\r^2} + {\half k \r^2}-\frac{GMm}{\sqrt{\r^2+z^2}}.
\eeq

\begin{figure}[t ]
\subfigure[ ~$l=0$ \label{fig4a}]{\includegraphics[width=4cm]{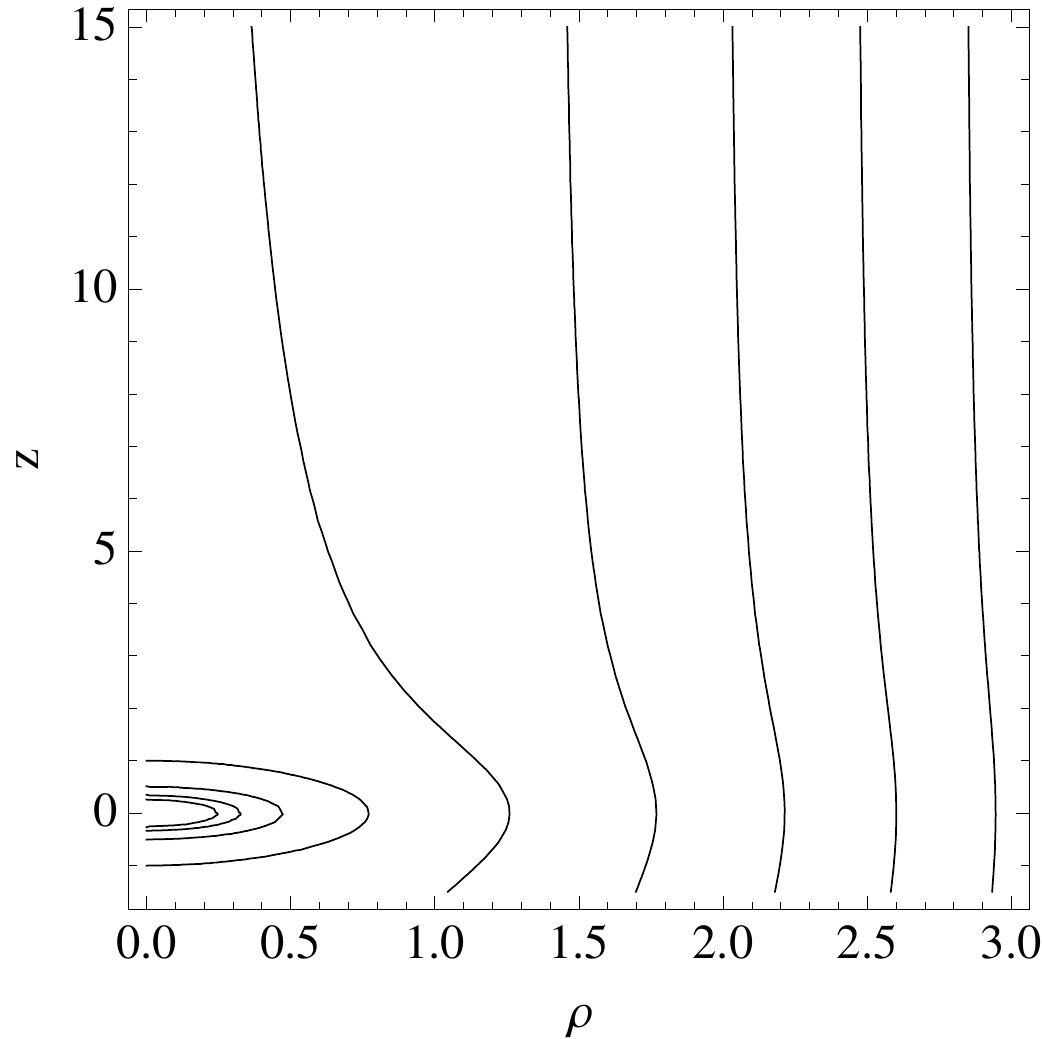}}
\subfigure[~$l=0.12$ \label{fig4b}]{\includegraphics[width=4cm]{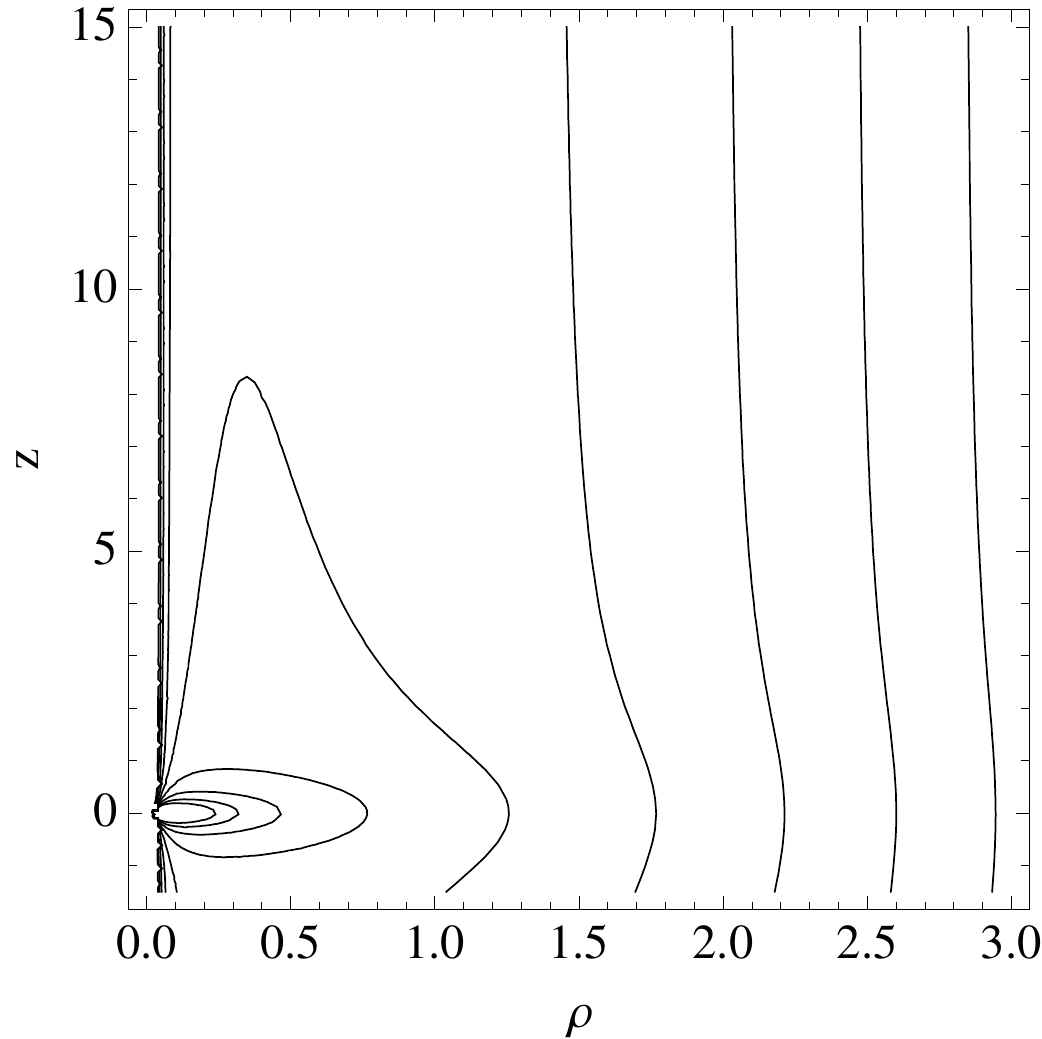}}\\
\subfigure[~$l=1.5$ \label{fig4c}]{\includegraphics[width=4cm]{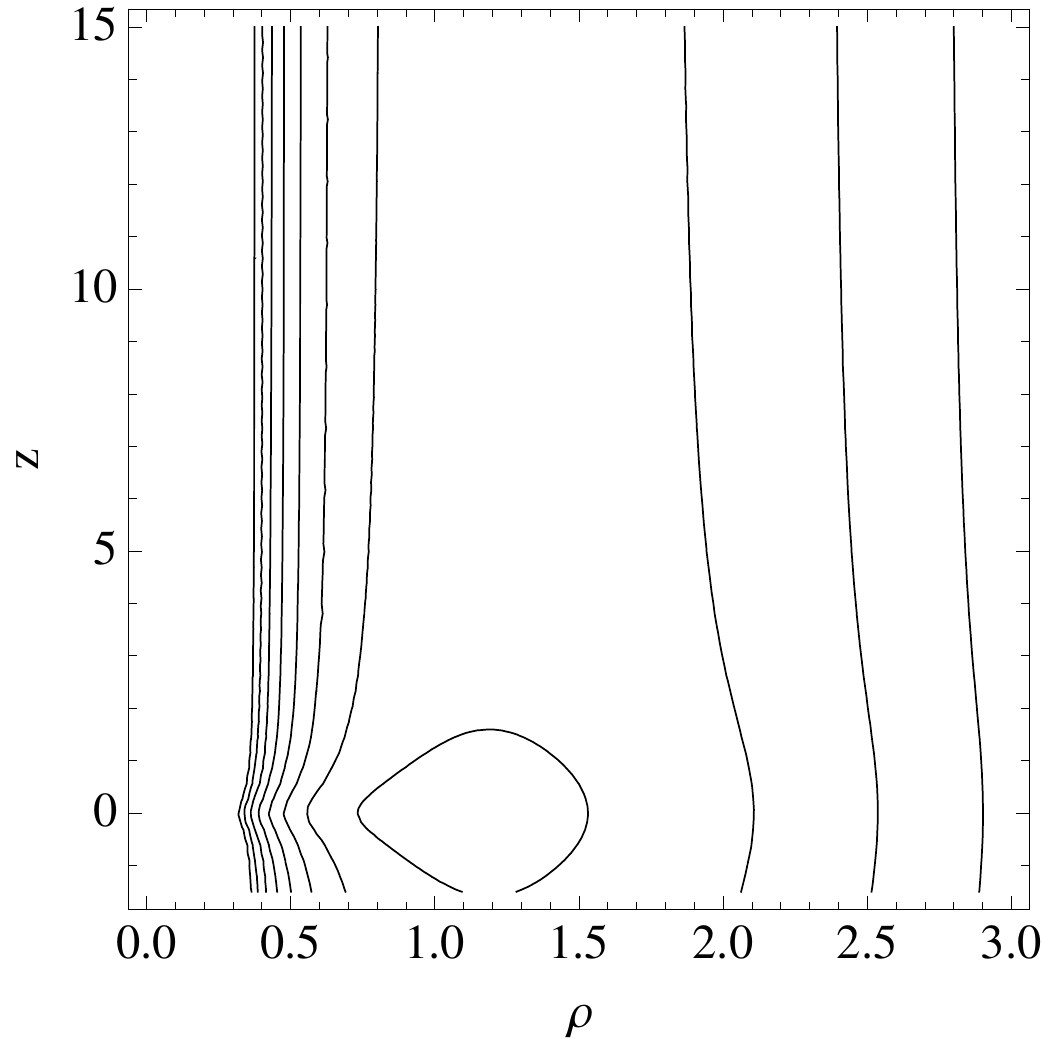}}
\subfigure[~$l=2$ \label{fig4d}]{\includegraphics[width=4cm]{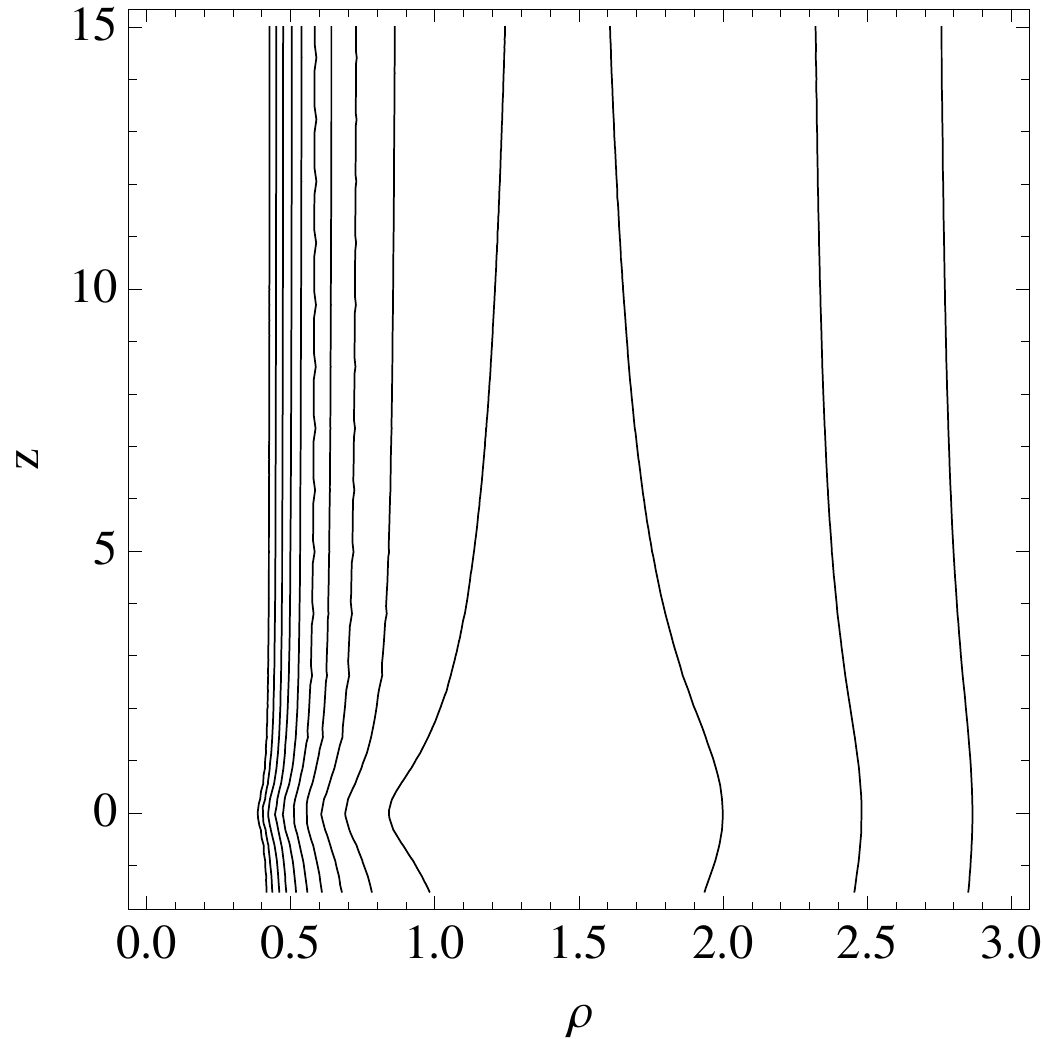}}
\caption{\label{fig4}  Newtonian elastic spinning ring of total mass $m$, moving axisymmetrically in the gravitational field of a point mass $M$. Contour plots of the boundaries on the motion for different energies and for angular momentum 
$l=0,\, 0.12,\, 1.5,\, 2$ (in units with $m=k=GM=1$). Both bound 
trajectories and escape trajectories can be present, 
depending on the energy.}
\end{figure}

The bounding contour is given in Fig.~\ref{fig4} for 
different values of the angular momentum and energy, in units with $m=k=GM=1$.
The angular momentum term $\propto \r^{-2}$ dominates at small $\r$ and makes an
inner $\r$ wall,
while the elastic energy term $\propto\r^2$ dominates at large $\r$ 
and makes an outer $\r$.
The gravitational potential
well deforms the inner wall inward toward the center,
and the outer wall outward. 
The figure illustrates that,
depending on 
the energy and angular momentum, the ring can either oscillate in a confined toroidal 
region around the equatorial plane,  or escape to infinity confined to a channel 
along the azimuthal direction. In the exceptional case of
zero angular momentum, the ring can be confined to a spherical region
including the origin.

\section{String dynamics}
\label{motion}

\subsection{String in flat background}
\label{flatbackground}

In order to demonstrate the effect of the current on the motion of the string 
in a simplified setting, 
and to obtain a useful formula for the string energy far from the black hole,
let us first consider the case in which the background is just flat spacetime.
We use cylindrical coordinates $(t,\rho,z,\phi)$, so $g_{\mu\nu}={\rm diag}(-1,1,1,\rho^2)$, and we consider
axisymmetric motion. 
The effective potential (\ref{VKerr}) then takes the form
\beq\label{Vflat}
V(\r)=-E^2 + \r^2(\tS^{\t\t})^2.
\eeq
The motion is confined to the region where $V(\r)\le0$. 

The bounds of the motion are defined by 
$V(\r)=0$, i.e.\ 
\beq\label{V0flat}
E=E_b=\r \tS^{\t\t}=\mu\r +J^2/\r.
\eeq
The right hand side is $1/2\pi$ times 
the Killing energy
of the string when $\dot{r}=\dot{\theta}=0$.
Since the string is then at rest, and there is no
redshift factor in flat spacetime, this is the same as
the energy measured in the rest frame of the string.
The first term represents the 
elastic energy of the string, while the 
second term is 
the angular momentum barrier arising 
from the current circulating in both
directions around the string. Alternatively, it 
can be viewed as a 
variable ``rest mass" 
of the string, arising from the
energy density $J^2/\r^2$ of the 
current.
In the presence of the current, the string 
oscillates between two 
$\r$ values, the roots of (\ref{V0flat}), 
never collapsing to a point.

The elastic energy and angular momentum
play similar roles in the relativistic case 
(\ref{V0flat}) and the Newtonian
case (\ref{Ncontour}), but with different powers
of $\r$. The relativistic elastic energy is proportional
to the string length rather than its square,
and the angular momentum barrier comes 
from a kinetic energy that is linear rather than 
quadratic in the momentum.
A more important difference between the Newtonian
elastic ring and the relativistic string is that 
in the latter the current can
circulate in either direction. Both directions can
simultaneously contribute to the angular momentum barrier, so the
barrier is not determined by the net angular momentum.
Indeed it is present even when the net angular momentum
vanishes, which is the case when either $j_\s$ or $j_\t$ vanishes.

\subsection{String in Kerr background}

\subsubsection{General setting}

Next we consider the full problem, 
{\em i.e.}\ the motion of the string in Kerr spacetime, with line element 
(in Boyer-Lindquist coordinates)
\begin{align}\label{Kerr}
ds^2=&-\left(1-\frac{2Mr}{R^2}\right) dt^2
-\frac{4M r a\sin^2\theta}{R^2} dt d\phi\nonumber\\
&\quad+\left(r^2+a^2+\frac{2M r a^2}{R^2} \sin^2\theta\right)\sin^2\theta d \phi^2\nonumber\\
&\qquad +\frac{R^2}{\Delta^2} dr^2
+R^2 d\theta^2.
\end{align}
Here $M$ and $a=J/M$ are the mass and the specific 
angular momentum of the black hole respectively, 
\beq
R^2=r^2+a^2\cos^2\theta, \quad \Delta^2=r^2-2M r+a^2,
\eeq 
and we are using units in which $G=c=1$.
(Below we shall instead choose units with $GM=1$.)
The event horizon is located where $\D=0$, at 
\beq
r_h=M+\sqrt{M^2-a^2},
\eeq
and the boundary of the ergoregion lies where the coefficient of $dt^2$ vanishes, at
\beq 
r_e=M+\sqrt{M^2-a^2 \cos^2\theta}.
\eeq
We consider only motion outside the horizon, so 
this coordinate system is adequate.

\begin{figure}[t]
\includegraphics[width=8cm]{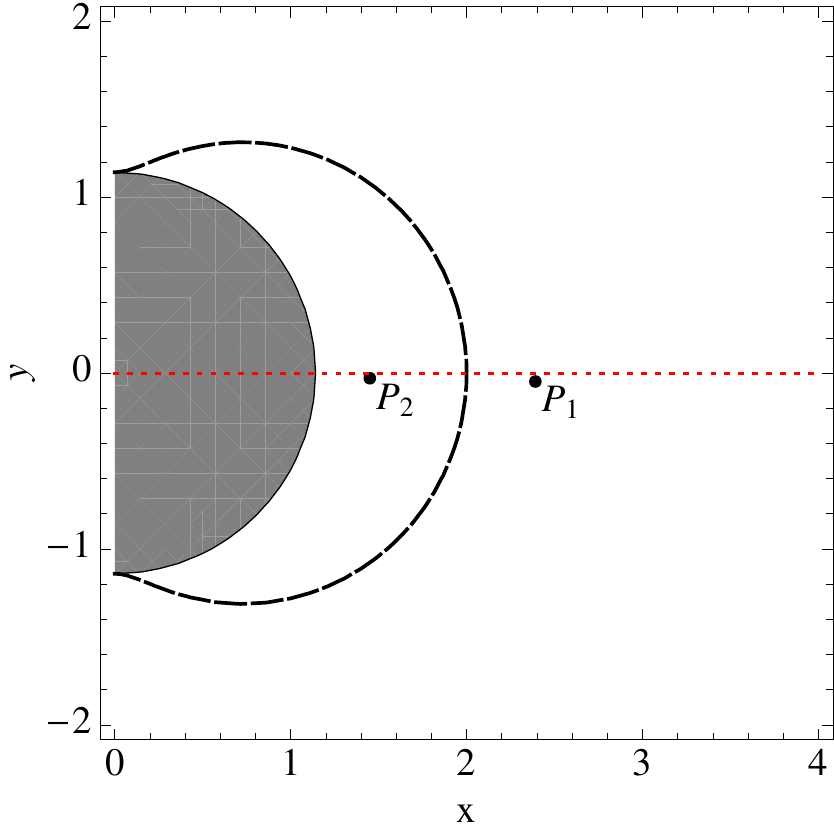}
\caption{\label{figinit} Graphical representation of the initial conditions, in units with $GM/c^2=1$. 
The area inside the horizon is covered in grey, the dotted red line marks the equatorial plane and the dashed 
black curve is the boundary of the ergoregion. $P_1$ corresponds to $r=r_h+1.25$ and $P_2$ to $r=r_{\rm ISCO}$. 
At both points, $\theta=\pi/2+0.02$, and we choose $\dot{r}=\dot{\theta}=0$. The specific angular momentum
is $a=0.99$ and, therefore,  $P_2$ lies within the ergoregion ($r_{\rm ISCO} < r_e$).}
\end{figure}

The explicit form of the effective potential (\ref{VKerr})
is too complicated to be illuminating, unlike the flat background case.
A given orbit is determined once  $j_s$, $j_t$, and 
initial conditions are specified. 
The initial conditions determine the values of the energy 
$E$ and angular momentum $L$, 
and the
subsequent motion is confined to the region contained in
the contour (\ref{VKerr0}).
The equations of motion can be integrated numerically
to explore the detailed behavior of the string motion.

We will present here only results for the string motion around a rapidly spinning
black hole, with $a=0.99$. 
This is complementary to the $a=0$ case presented by Larsen~\cite{Larsen:1993nt}.
The role of the black hole spin is addressed below in subsection \ref{spin}.
We consider
 just two sets of initial conditions for several different values of $J^2$ and $\o$, 
 which appear to 
 be adequate for understanding the features of the motion. For both sets 
 the string starts out at rest, $\dot{r}_0=\dot{\theta}_0=0$.
 The initial polar angle is  fixed in both cases at
 $\theta_0=\pi/2+0.02$, just below the equatorial plane. 
 The two cases differ only in the initial value of $r$. 
 The two values we will explore are (somewhat arbitrarily) 
 $r_0=r_h+1.25 M\approx 2.39\, M$ and $r_0=r_{\rm ISCO}\approx 1.45\, M $,
where
$r_{\rm ISCO}$ is the radius of the innermost stable co-rotating circular orbit (ISCO) 
which is given in terms of $M$ and $a$ in, e.g., Ref.~\cite{Bardeen:1972fi}.

In Fig.~\ref{figinit} the two sets initial conditions correspond to points $P_1$ and $P_2$ respectively. 
Each point in this figure 
other than those on the $y$ axis 
corresponds to a circular ring in space. 
The coordinates $x$ and $y$, used as labels in this and later 
graphs of the $(r,\theta)$ plane, are defined by
 \begin{align}
 & x=r \sin \theta,\\
 & y=r\cos\theta.
 \end{align}

The grey area indicates the region inside the horizon, the dotted red line marks the equatorial plane and the dashed 
black curve is the boundary of the ergoregion. For $\o$ we consider only the three values $\o=0, \pm1$,  since these
span the full dynamical range and can be expected to frame the full picture. 
Note that changing the sign of $\omega$ is equivalent to changing the sign of $a$, since 
$\o$ appears always multiplied by $a$ (via $g_{t\phi}$)  
in the equations of motion, and all other occurrences of $a$ are via $a^2$. 
Positive $\o$ corresponds to co-rotation, while negative 
$\o$ corresponds to counter-rotation. 

\subsubsection{Boundaries of motion}
\label{boundaries}

By studying the boundaries of the motion we can classify the types of dynamics and determine how
they depend on the current, energy, and initial conditions. These boundaries are determined
by eq.~(\ref{VKerr0}). For a given initial condition and current, a particular boundary is determined.
These boundaries 
are the relativistic version of the contours plotted in Fig.~\ref{fig4} for a 
Newtonian elastic spinning ring and, as we will see shortly, the share many qualitative characteristics 
with their Newtonian counterparts. 
At the end of this subsection we give examples of such boundaries, but first we present some graphs, 
deduced from the potential, that allow one to see which ranges of current or energy will correspond 
to different qualitative types of boundaries.

\begin{figure}[t]
\subfigure[  $~\omega=1$ \label{Ebwm1}]{\includegraphics[width=6.5cm]{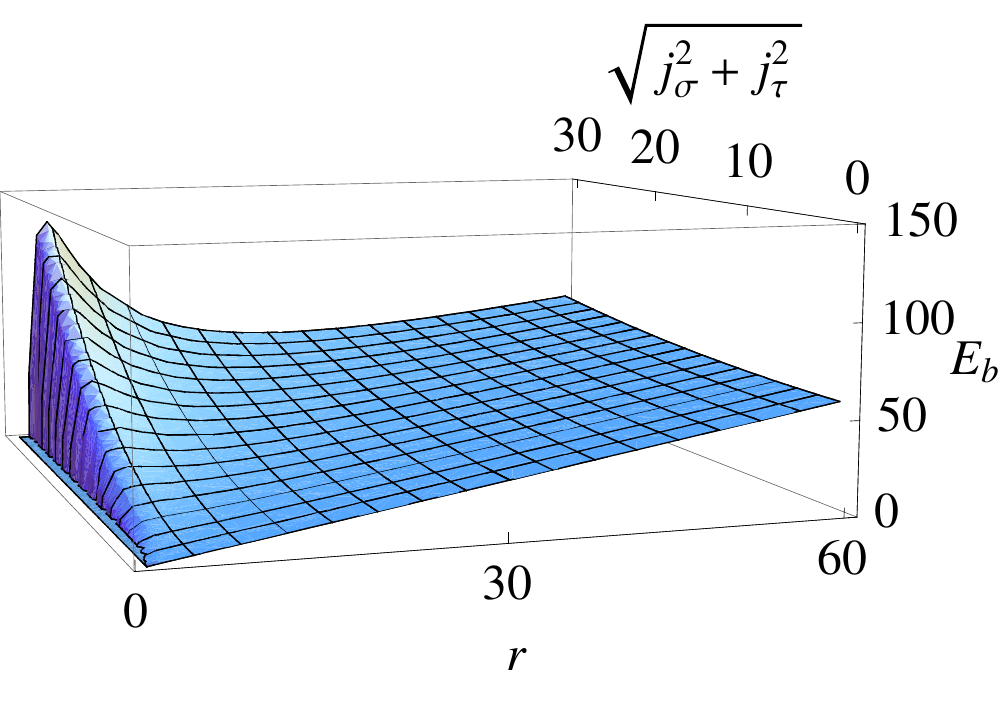}}\\
\subfigure[$~\omega=0$ \label{Ebw0}]{\includegraphics[width=6.5cm]{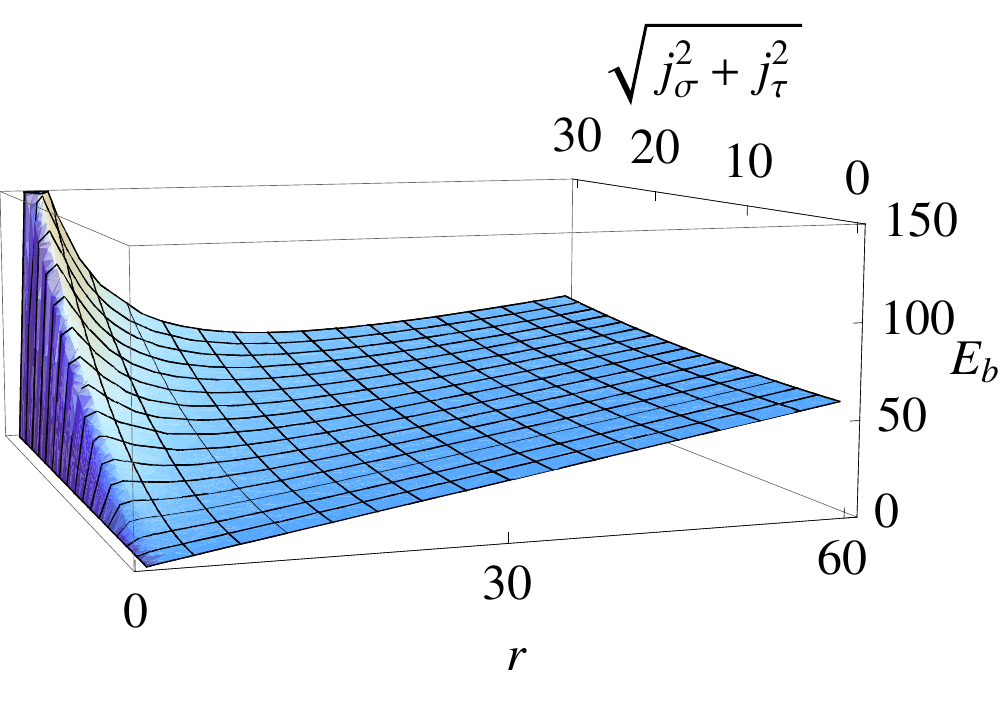}}\\
\subfigure[$~\omega=-1$ \label{Ebw1}]{\includegraphics[width=6.5cm]{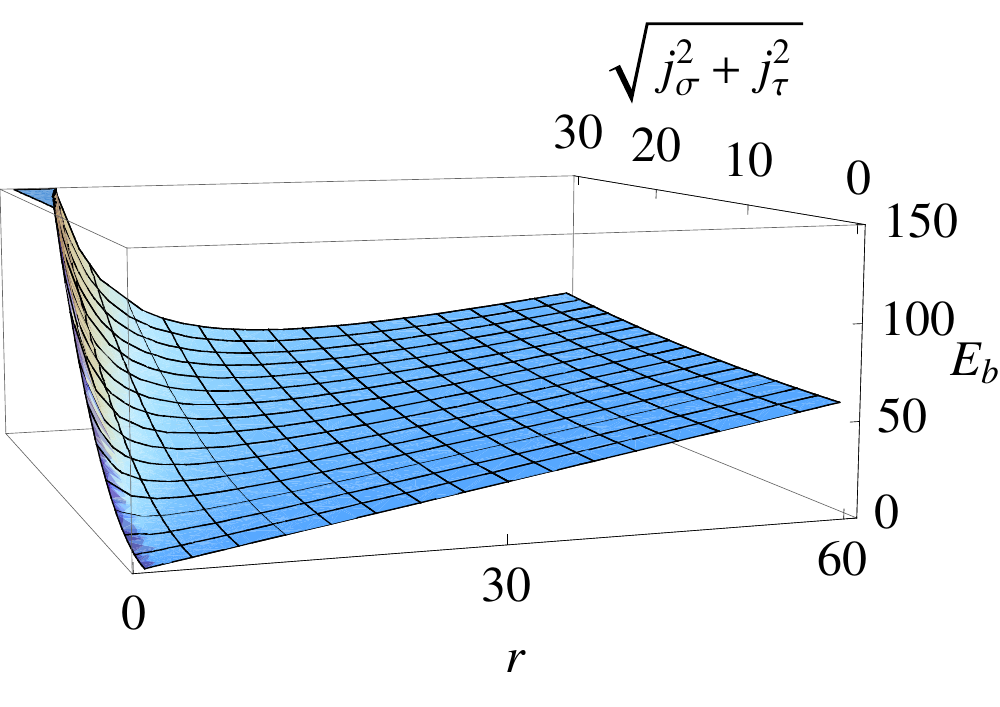}}
\caption{\label{Ebfig}  Plots of the energy of a string at rest (or energy at the boundary of motion) as a function of $r$ (plots starting at $r=r_h$) and  
$\sqrt{j_\s^2+j_\t^2}$,
for three values of the ratio between the components of the current $\omega=-j_\s/j_\t$ and at $\theta=\pi/2+0.02$ ($GM=c=\mu=1$). 
All graphs possess
minima, which implies that bound motion is possible for energies 
greater than $E_b^{\min}$, the energy 
associated with the minimum. However, only for $\omega=0,-1$ 
is there a maximum located outside the horizon.
}
\end{figure} 
\begin{figure*}[t]
\subfigure[ ~$r=r_h+1.25$, $\omega=1$ \label{Ejqswm1}]{\includegraphics[width=5.5cm]{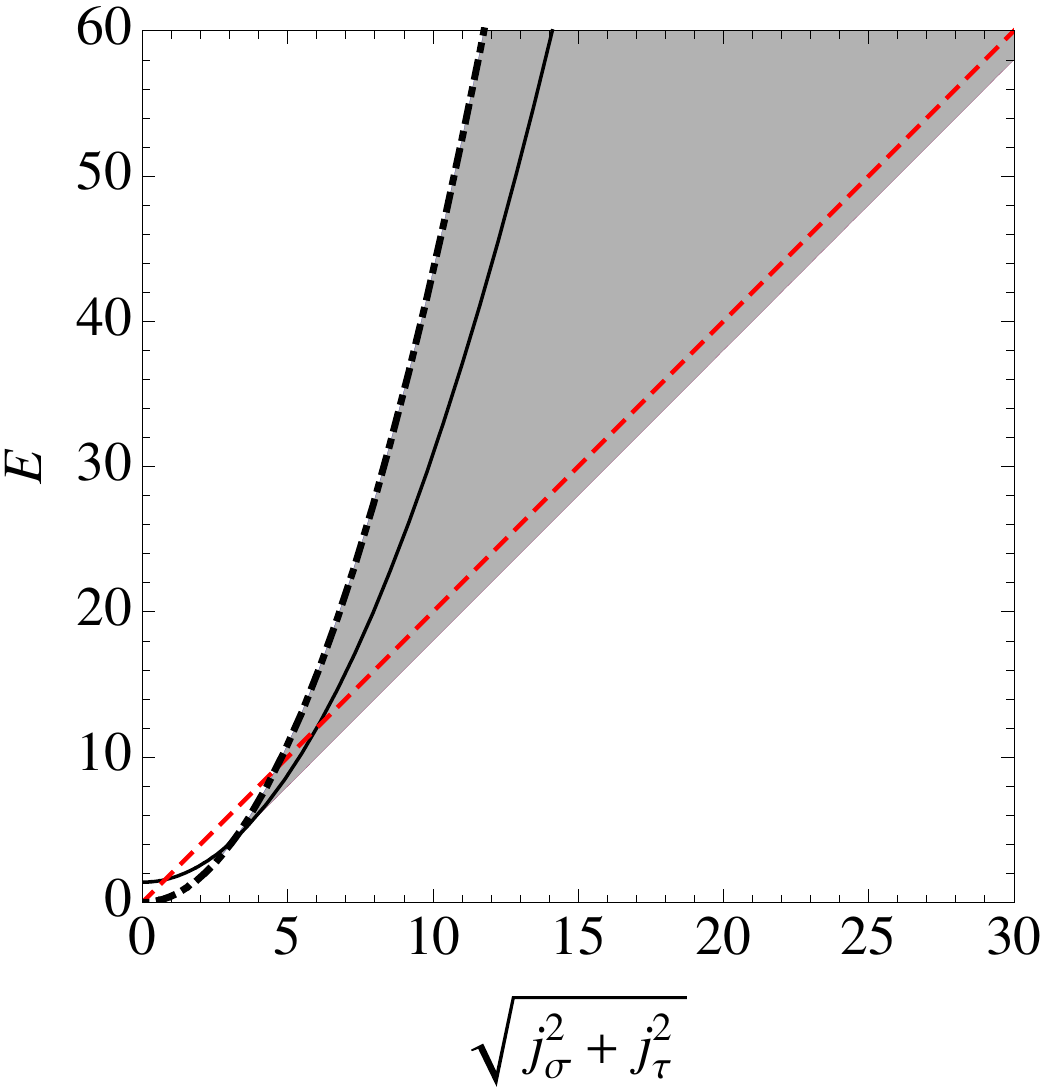}}
\subfigure[~$r=r_h+1.25$, $\omega=0$ \label{Ejqsw0}]{\includegraphics[width=5.5cm]{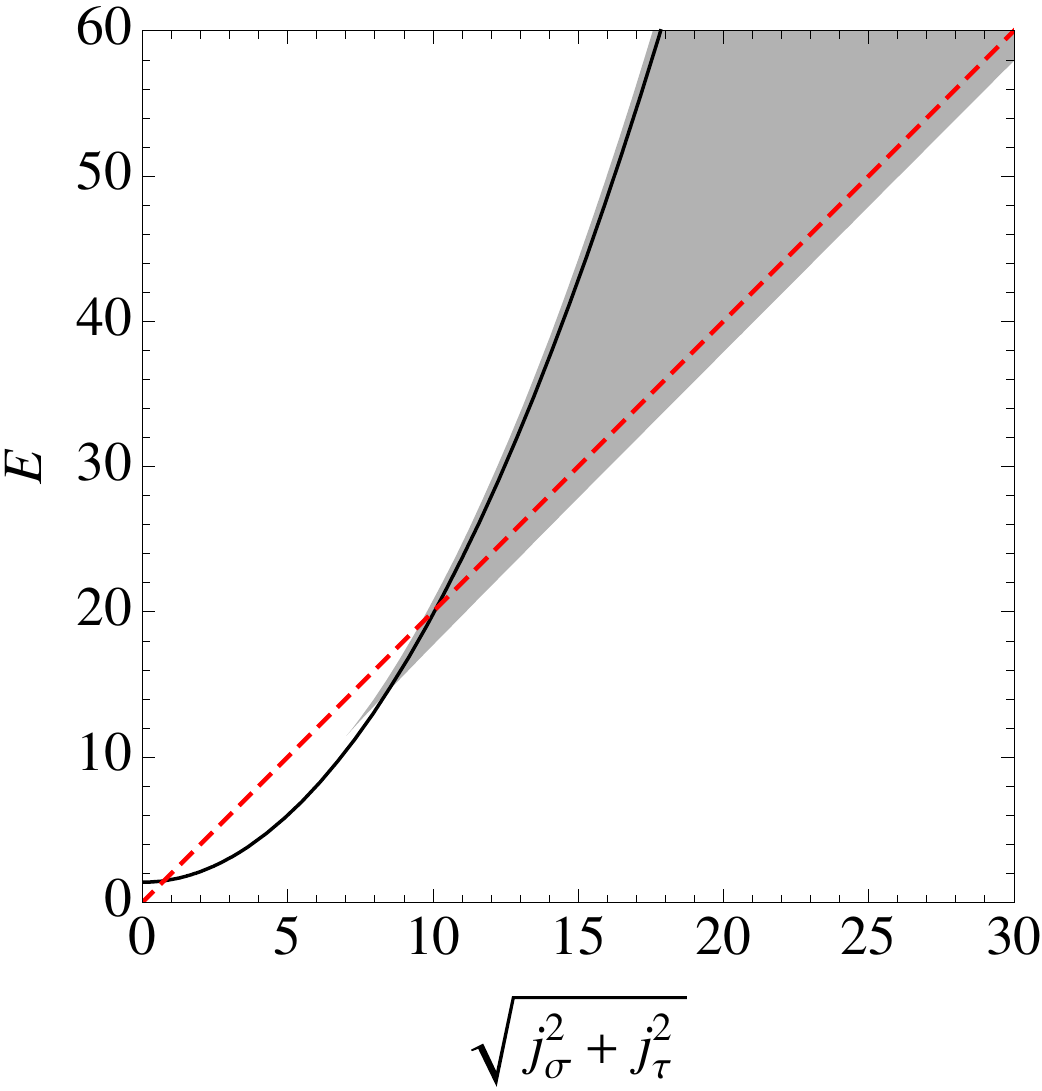}}
\subfigure[~$r=r_h+1.25$, $\omega=-1$ \label{Ejqsw1}]{\includegraphics[width=5.5cm]{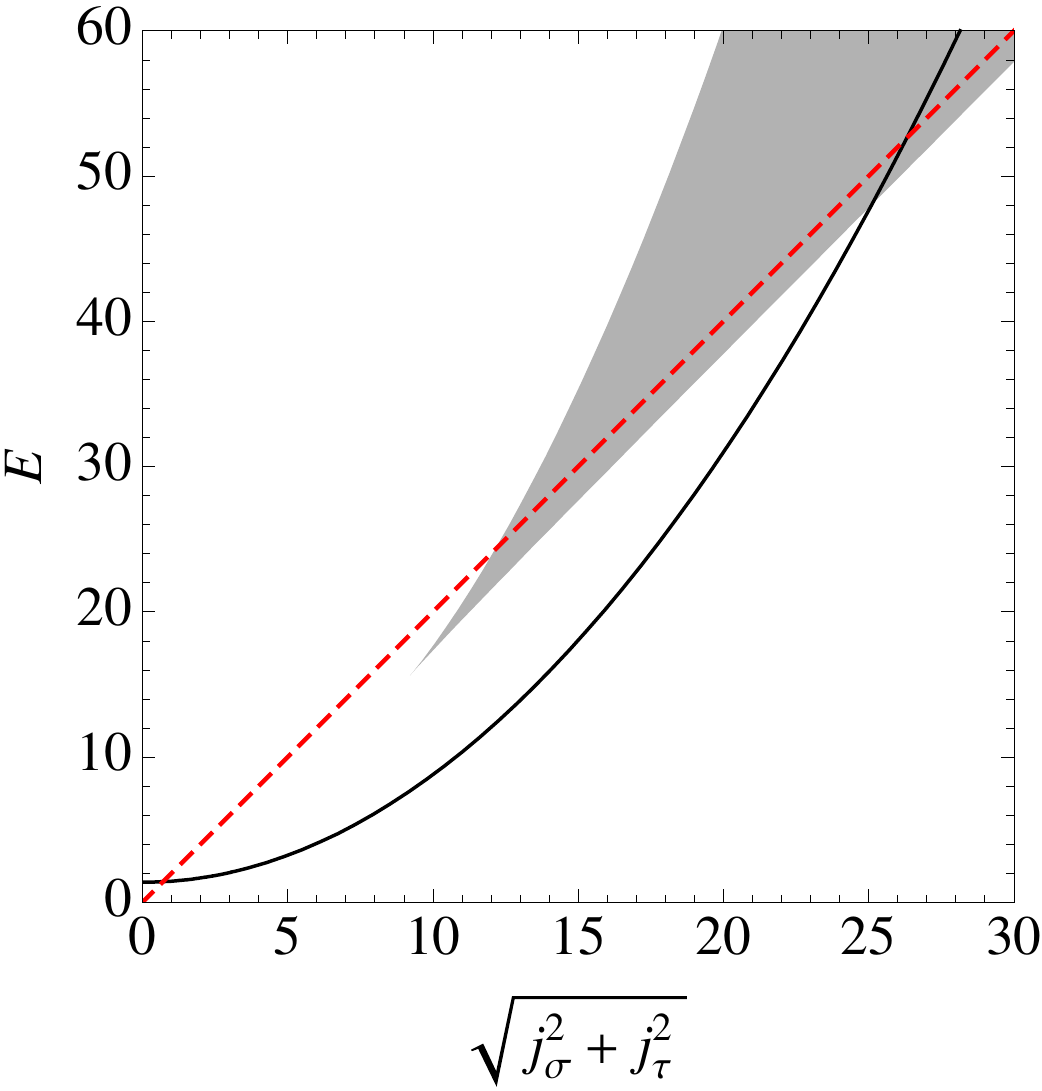}}\\
\subfigure[ ~$r=r_{\rm ISCO}$, $\omega=1$ \label{Ejqswm1in}]{\includegraphics[width=5.5cm]{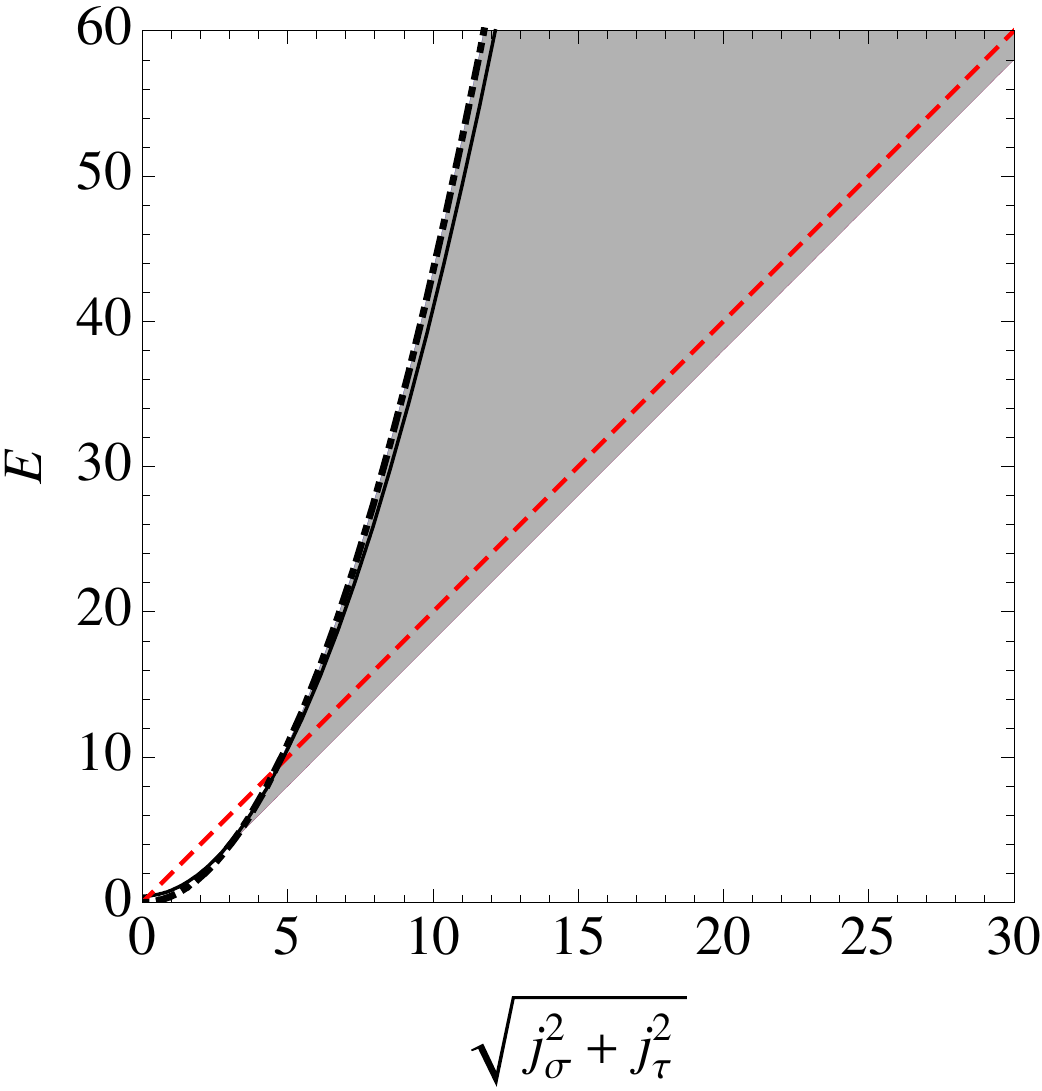}}
\subfigure[~$r=r_{\rm ISCO}$, $\omega=0$ \label{Ejqsw0in}]{\includegraphics[width=5.5cm]{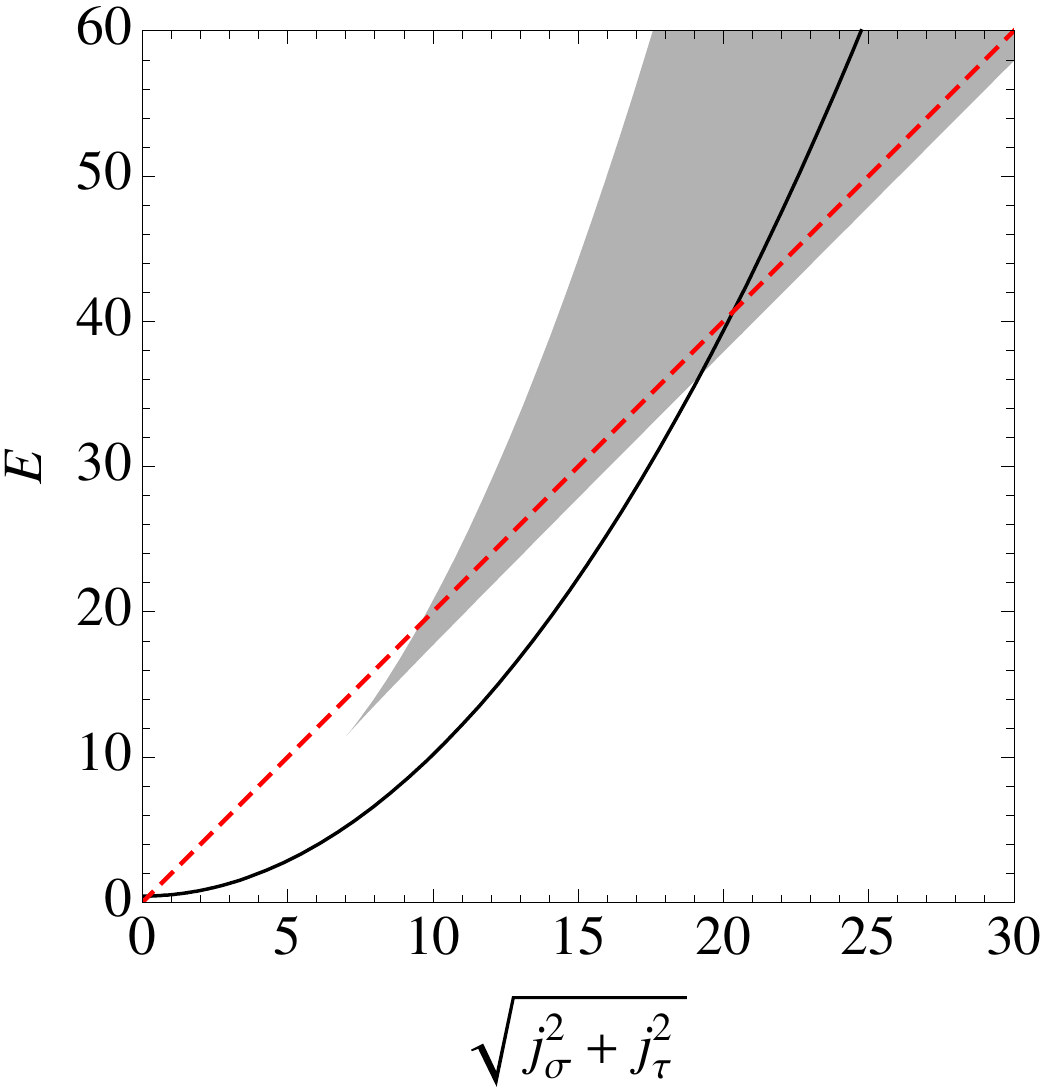}}
\subfigure[~$r=r_{\rm ISCO}$, $\omega=-1$ \label{Ejqsw1in}]{\includegraphics[width=5.5cm]{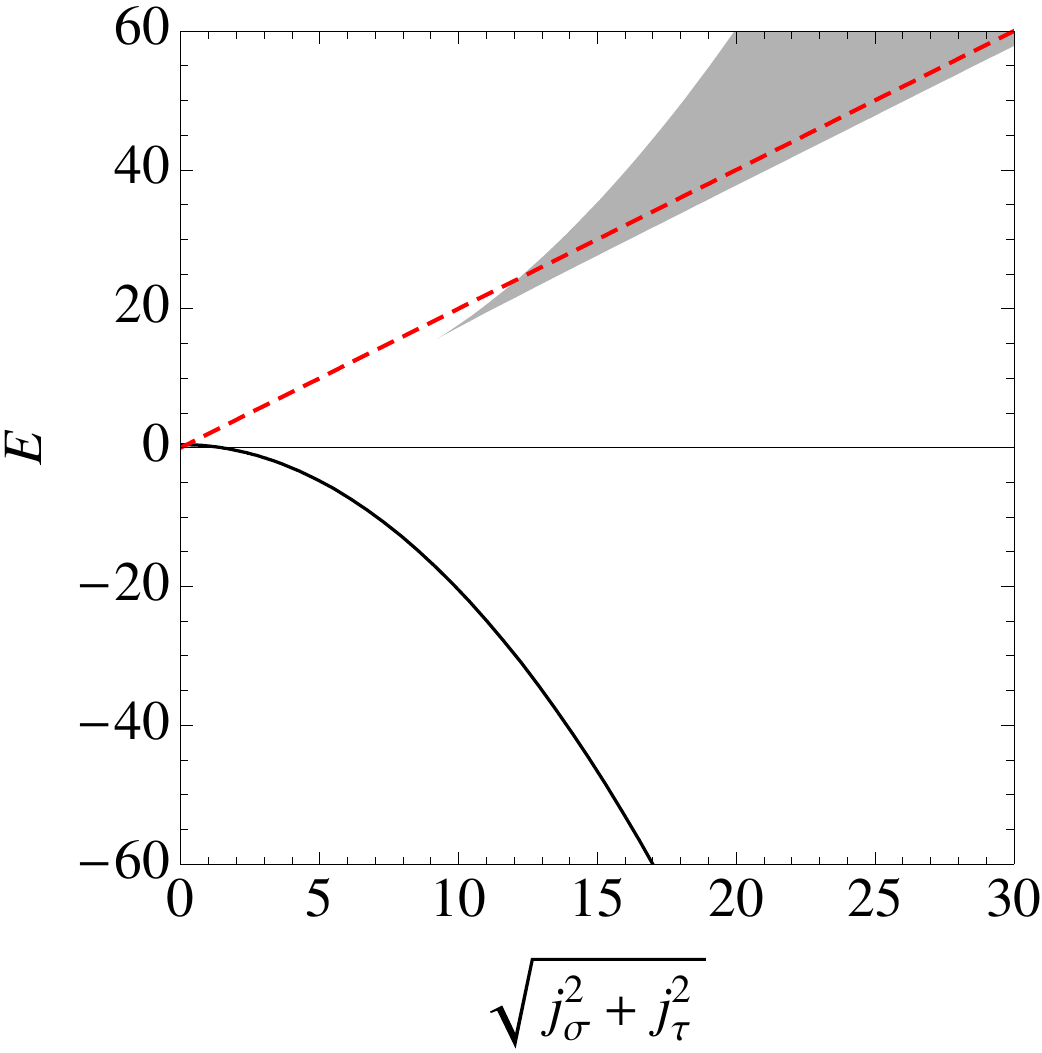}}
\caption{\label{Ejsq}  Plots of the energy of a string at rest (or at the boundary of motion) as a function of 
$J=\sqrt{j_\s^2+j_t^2}$
for specific values of  $\omega$ and $r$ as indicated, and for $\theta=\pi/2+0.02$ ($GM=c=\mu=1$).  
The solid black curve,
which can be thought of as a constant $r$ slice of the graphs in Fig.~\ref{Ebfig},
is the energy of the string at rest at the given radius $r$, and is the only 
curve on the graph that depends on the value of $r$. 
The dashed red line corresponds to the energy of a string at rest at infinity (the escape energy). 
The shaded region is the area between the curves corresponding to the minimum and the 
maximum of $E_b$ for each $J$  
(cf.~Fig.~\ref{Ebfig}), with the exception of the $\omega=1$ 
cases where the dotted-dashed black curve corresponds to the value of $E_b$ at the horizon. 
These graphs  
allow one to infer
the properties of the motion.
For a given energy or value of $J$, there is a corresponding point on the solid black curve. 
If this point lies 
outside the shaded area and above the dashed red line then the motion has just an outer boundary
in the $x$ direction,  but 
the string has enough energy to potentially escape
(there is no boundary in the $y$ direction). 
If the point lies
outside the shaded area and 
below the dashed red line then the motion 
is bounded in all directions. 
If instead the point lies 
within the shaded region, then the motion has boundaries both at low and high values of 
$x$.
Additionally, if the point lies above the dashed red line then there is no boundary on the 
$y$ motion
and the string can potentially escape. If the point
lies below the dashed red line then 
the motion 
can be trapped in some toroidal volume. 
Notice that for $\omega=-1$ and inside the ergosphere, 
Fig.~\ref{Ejqsw1in}, there can never be an inner boundary.} 
\end{figure*}

To begin with, 
in Fig.~\ref{Ebfig} we have plotted $E_b$ as a function of 
$J=\sqrt{j_\s^2+j_\t^2}$ and $r$ for fixed $\theta=\pi/2+0.02$, 
which corresponds to our initial conditions, and for three different values of $\omega$. 
In this and all subsequent plots we adopt 
units with $GM=c=\mu=1$. That is, in terms of the gravitational radius
$r_g=GM/c^2$, what is plotted
is $E_b/\mu r_g$ and $J\sqrt{c/\mu}/r_g$ vs. $r/r_g$.

In the cases $\o=0,-1$, where there is  
a maximum outside the horizon for any
given value of $J$, 
the motion has an outer boundary in $r$ but no inner boundary
if either $E<E_b^{\rm min}$ or $E>E_b^{\rm max}$, 
where $E_b^{\rm min}$ and $E_b^{\rm max}$ correspond to the minimum and the maximum of $E_b$ respectively. If $E_b^{\rm max}>E>E_b^{\rm min}$ then there will be three solutions to eq.~(\ref{VKerr0}), corresponding to an inner and an outer boundary and yet another outer boundary at low values of $r$.  
In the co-rotating case $\o=1$, where there is no maximum, 
the role of $E_b^{\rm max}$ is played by $E_b^h$, 
the value of $E_b$ at $r=r_h$. For both an inner and an outer boundary to exist, the energy must
fall in the range $E_b^{h}>E>E_b^{\rm min}$. Notice that 
in this case no string can exist outside the horizon with energy $E<E_b^{\rm min}$.

So far we have been discussing bounds in the radial direction near the equatorial plane.
There is also a boundary in the motion along the axis if the total energy of the string 
is less than the energy of a string at rest at infinity, since the string must have at least 
this much energy in order to be able to escape. We call this energy $E_{\rm esc}$.
To determine how it depends on $J$, we can use the 
flat background example of section \ref{flatbackground}. 
Eqn.~(\ref{V0flat})  gives $E_b=\mu\rho+J^2/\rho$, whose minimum lies at 
$E_{\rm esc}(J)=2J\sqrt{\mu}$.

 \begin{figure}[t ]
\subfigure[~$E=50$, $\omega=1$ \label{E50wm1}]{\includegraphics[width=4cm]{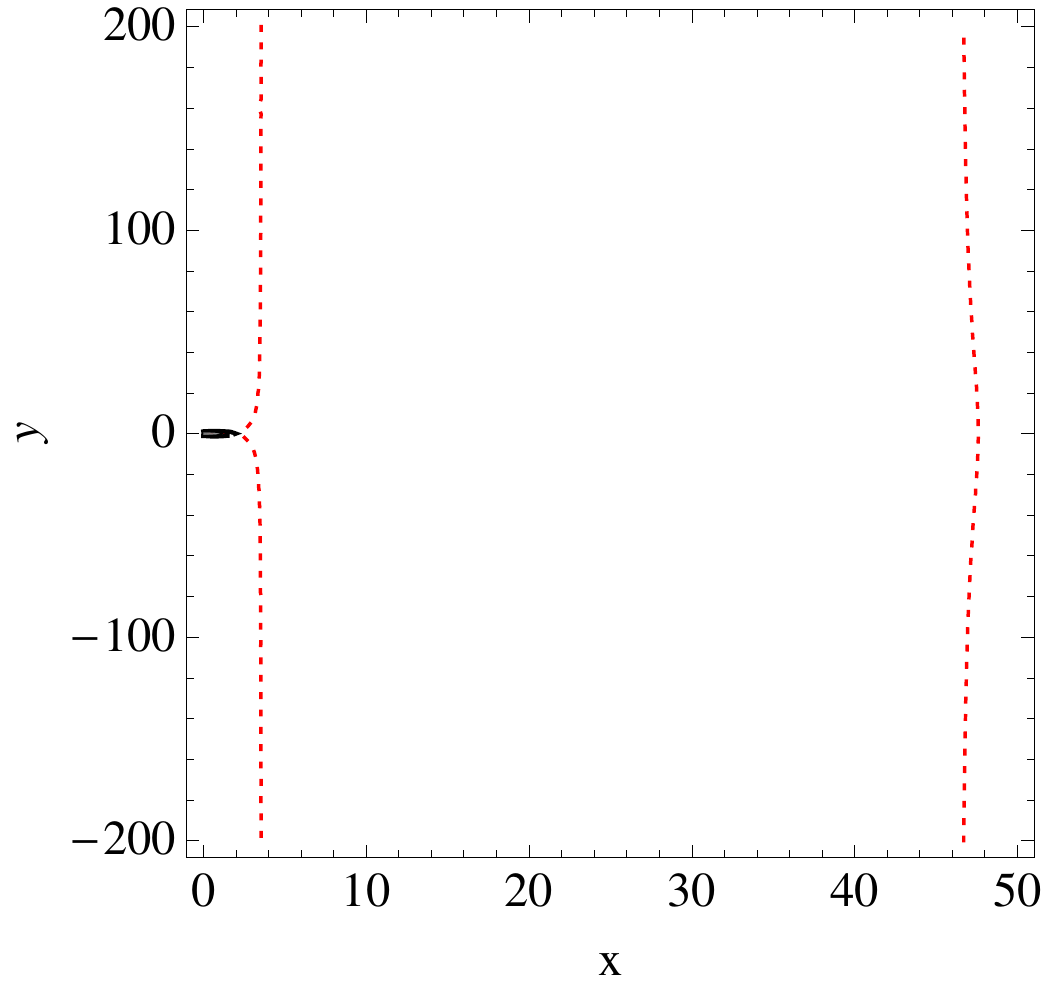}}
\subfigure[~$E=17$, $\omega=1$ \label{E17wm1}]{\includegraphics[width=4cm]{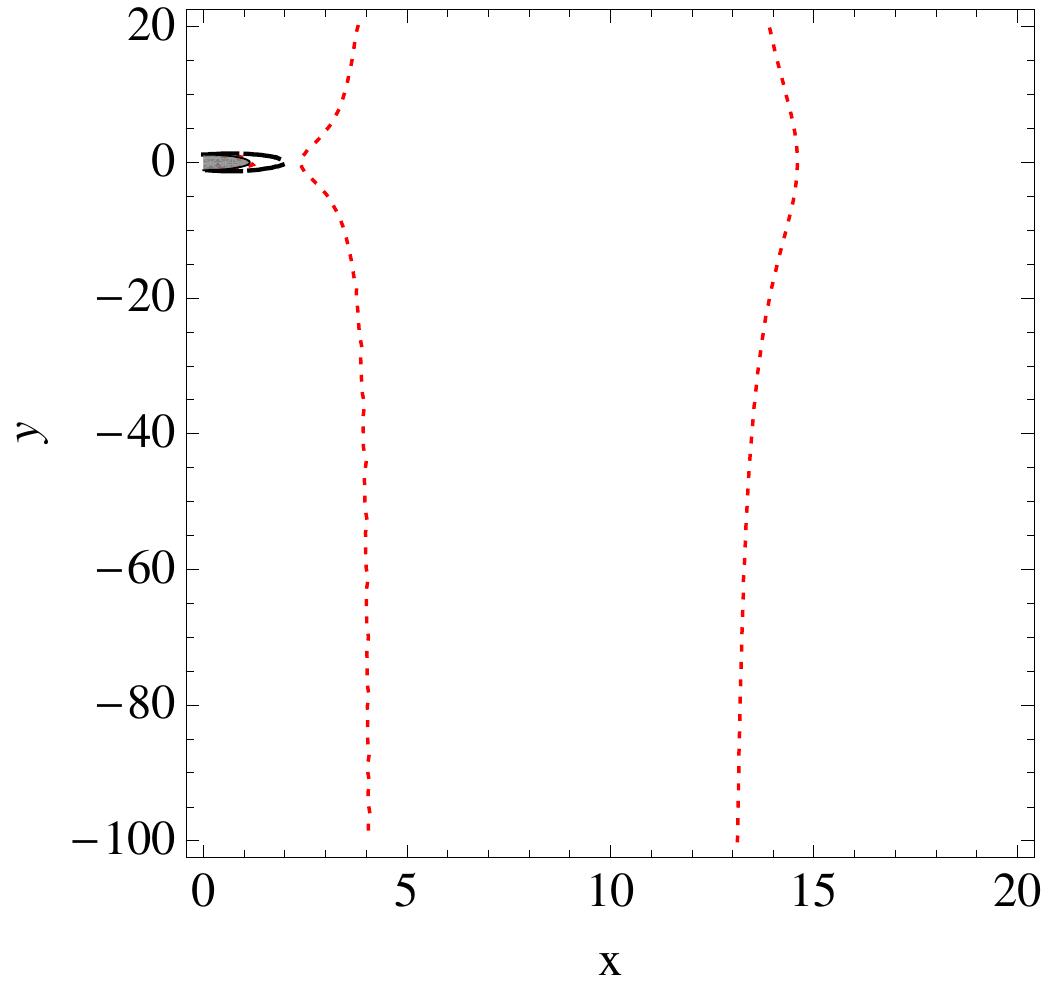}}\\
\subfigure[~$E=50$, $\omega=0$ \label{E50w0}]{\includegraphics[width=4cm]{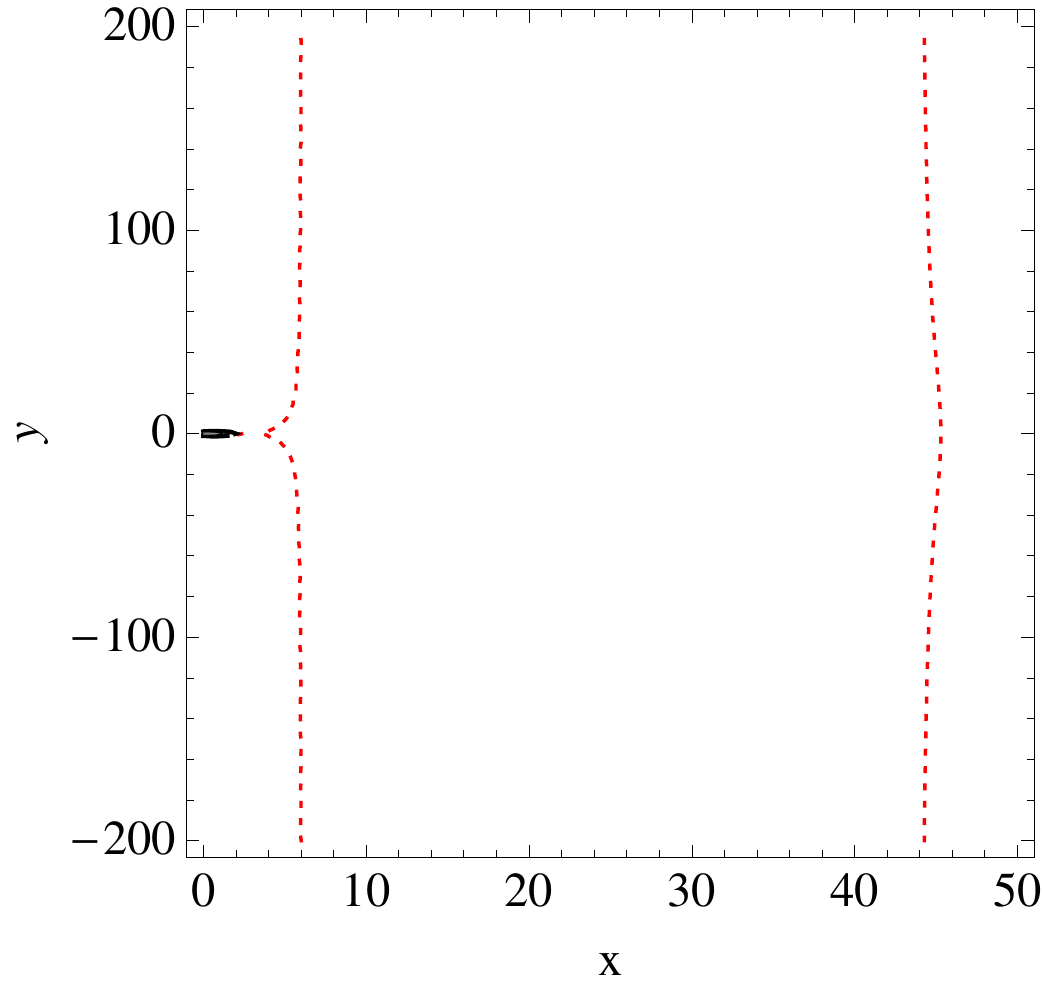}}
\subfigure[~$E=17$, $\omega=0$ \label{E17w0}]{\includegraphics[width=4cm]{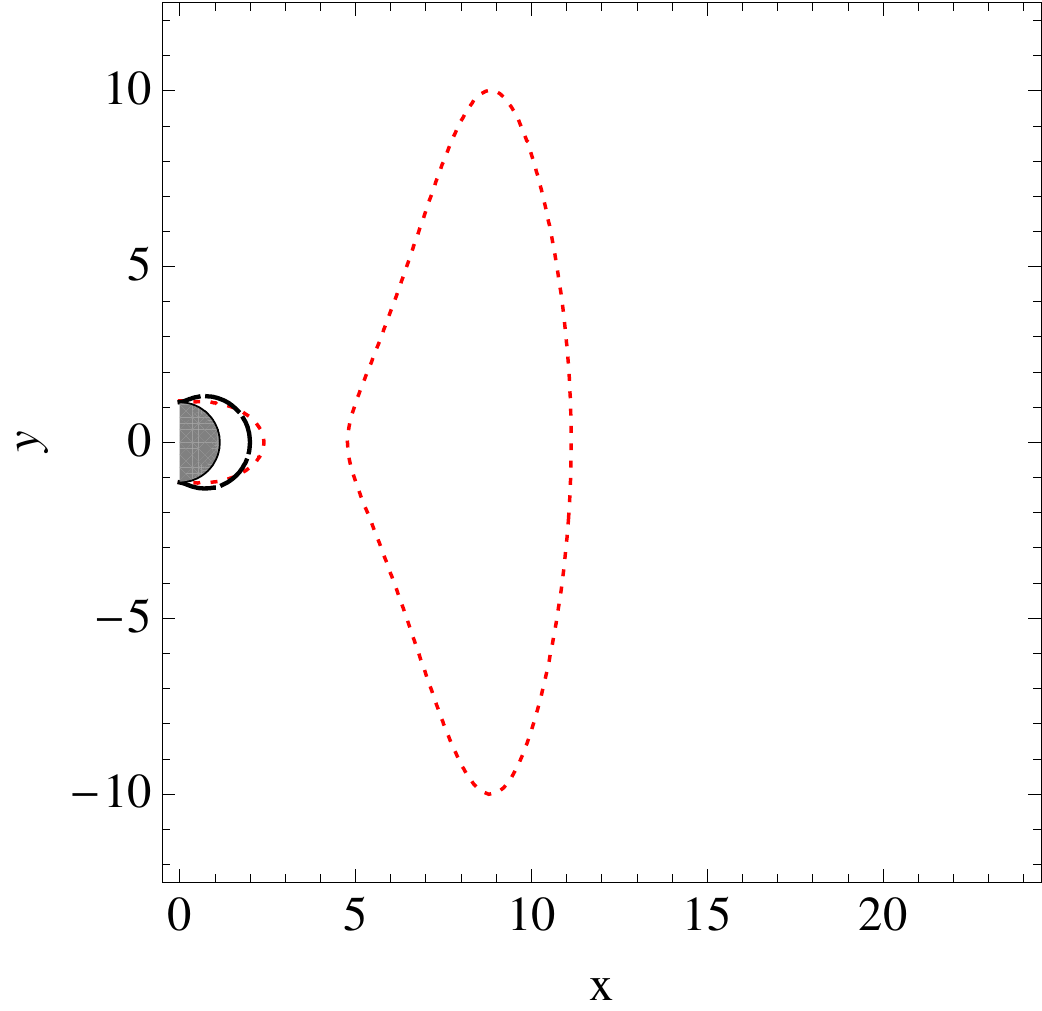}}\\
\subfigure[~$E=50$, $\omega=-1$ \label{E50w1}]{\includegraphics[width=4cm]{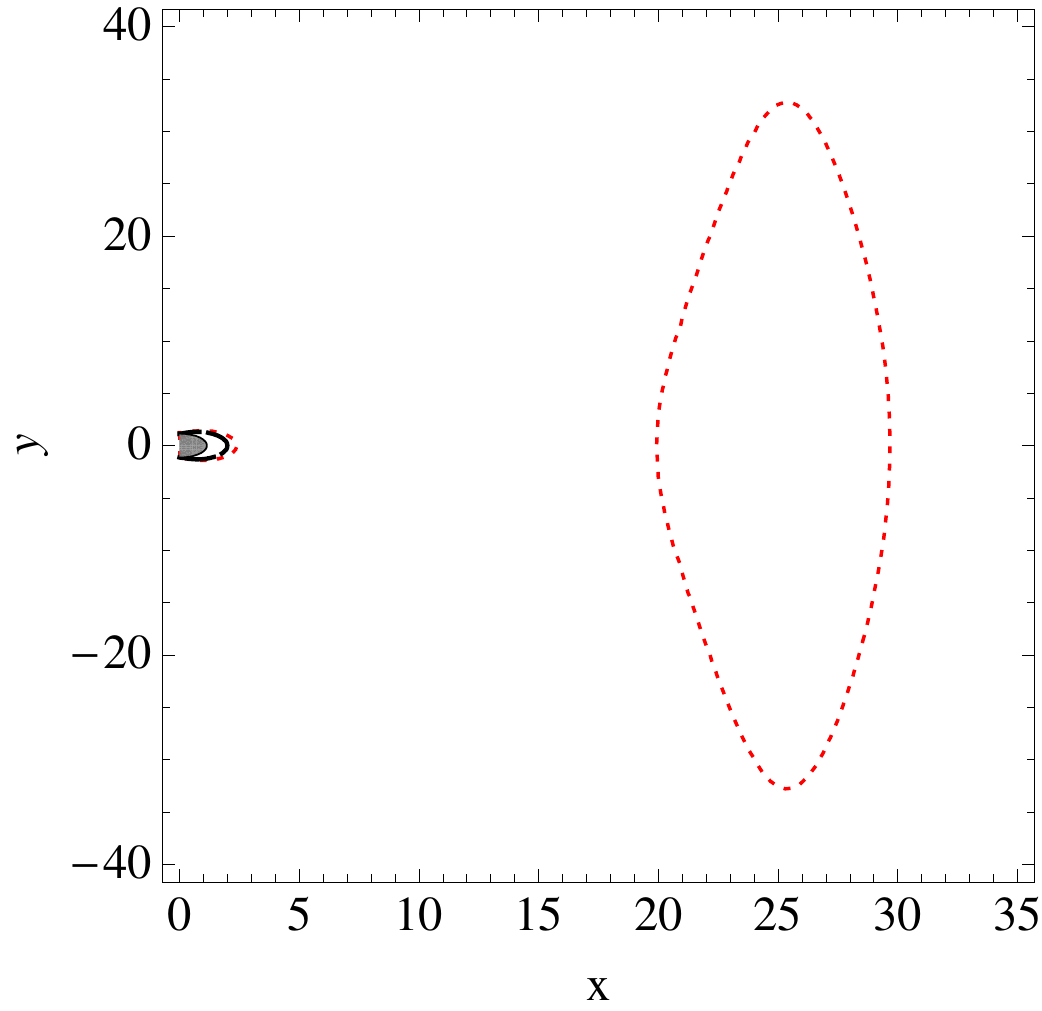}}
\subfigure[~$E=17$, $\omega=-1$ \label{E17w1}]{\includegraphics[width=4cm]{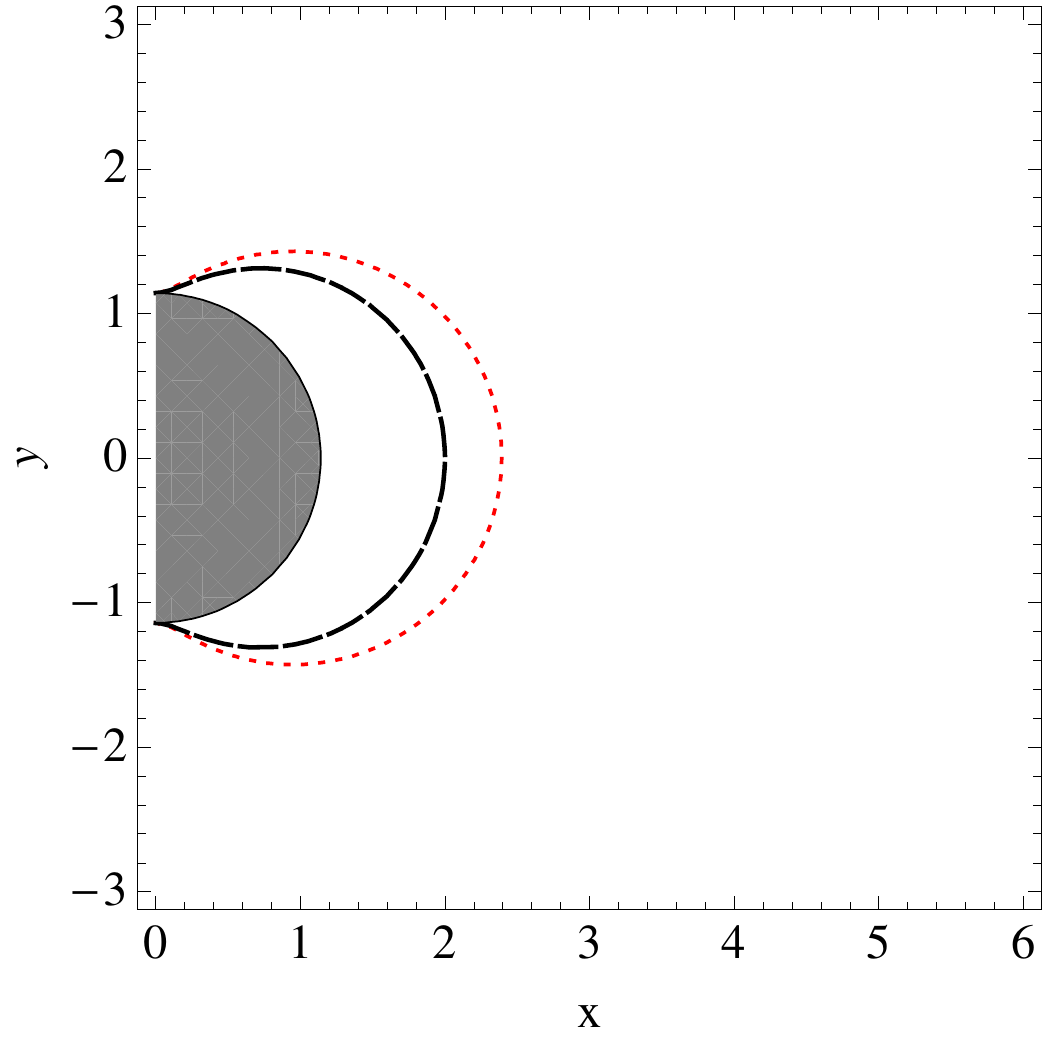}}
\caption{\label{Eoutfig}  Boundaries of motion for different values of the energy and 
the $\omega$ parameter, for  string starting at rest at $r=r_h+1.25$ and $\theta=\pi/2+0.02$. 
The grey area represent the region inside the event horizon, the dashed black curve the 
boundary of the ergoregion and the dashed red line the boundaries of motion. 
The form of the boundaries is exactly what one could have predicted by using 
Figs.~\ref{Ejqswm1}-\ref{Ejqsw1}.}
\end{figure}

Fig.~\ref{Ejsq} presents plots that allow one to see how the qualitative properties of the
motion depend on $J$, 
for specific values of  $\omega$ and the initial $r$ as indicated, 
and for $\theta=\pi/2+0.02$. 
The solid black curve,
which can be thought of as constant $r$ slice of the graphs in Fig.~\ref{Ebfig},
 is the energy (\ref{VKerr0}) of the string at rest at the given radius $r$, and is the only 
curve on the graph which depends on the value of the initial $r$.  
The dependence on $\o$ comes in only through the $g_{t\phi}$ term
in (\ref{VKerr0}). Increasing $\o$ acts to raise the 
energy for a given $J$.
The dashed red line corresponds to the energy of a string at rest at infinity (the escape energy). 
The shaded region is the area between the curves corresponding to the minimum and the 
maximum of $E_b$ for each $J$  
(cf.~Fig.~\ref{Ebfig}), with the exception of the $\omega=1$ cases where the 
dotted-dashed black curve corresponds to the value of $E_b$ at the 
horizon.\footnote{It is rather surprising to have a string (instantaneously)
at rest at the horizon 
with non-zero Killing energy. 
This comes about
as a singular limit whose nature is hidden by
a singular gauge. 
 The first term in eq.~(\ref{VKerr0}) vanishes on the horizon, so we have there
$E_b=-g_{t\phi}\tS^{\s\t}=2(g_{t\phi}/g_{\phi\phi})j_\s j_\t$, using eq.~(\ref{tSst}) in the last step.
The metric factors are finite and non-zero as long as the black hole is spinning, 
in which case 
the energy is non-zero as long as both $j_\s$ and $j_\t$
are non-zero. But if the string is instantaneously static
at the horizon, then the worldsheet is null on that slice, 
rather than being time-like, so the conformal gauge 
is not accessible. In the limit as such a configuration
is approached, the conformal gauge tangent 
vector $X^\mu_{,\tau}$ approaches the null horizon 
generator, becoming infinitely stretched since the 
gauge condition requires that it has
a nonzero timelike norm. 
Thus the current 4-vector
$j^\mu = X^\mu_{,\tau}j^\t + X^\mu_{,\s}j^\s$ itself diverges
as long as $j^\t=h^{\t\t}j_\t=g_{\phi\phi}^{-1} j_\t$ is non-zero.
A non-zero Killing energy configuration of an 
instantaneously static string at 
the horizon arises from this
singular limit of the current, but only, as seen above,
provided the black hole has spin. 
The underlying reason for this 
last requirement is that only in the presence of spin 
is the Killing vector $\partial_t$, 
with respect to which the energy is defined,
distinct from the null Killing vector that is normal to the
horizon.}

\begin{figure}[t ]
\subfigure[~$E=50$, $\omega=0$ \label{E50w0zoom}]{\includegraphics[width=4cm]{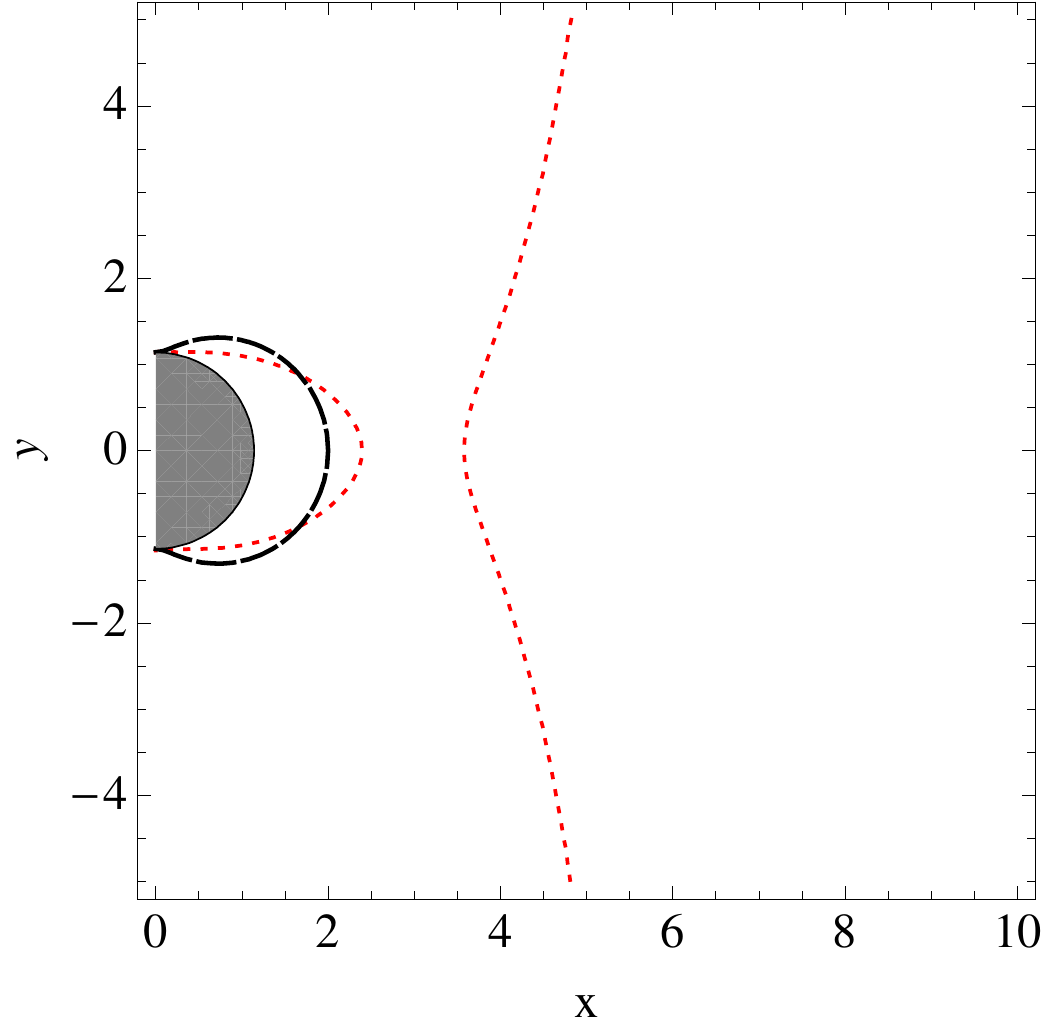}}
\subfigure[~$E=50$, $\omega=-1$ \label{E50w1zoom}]{\includegraphics[width=4cm]{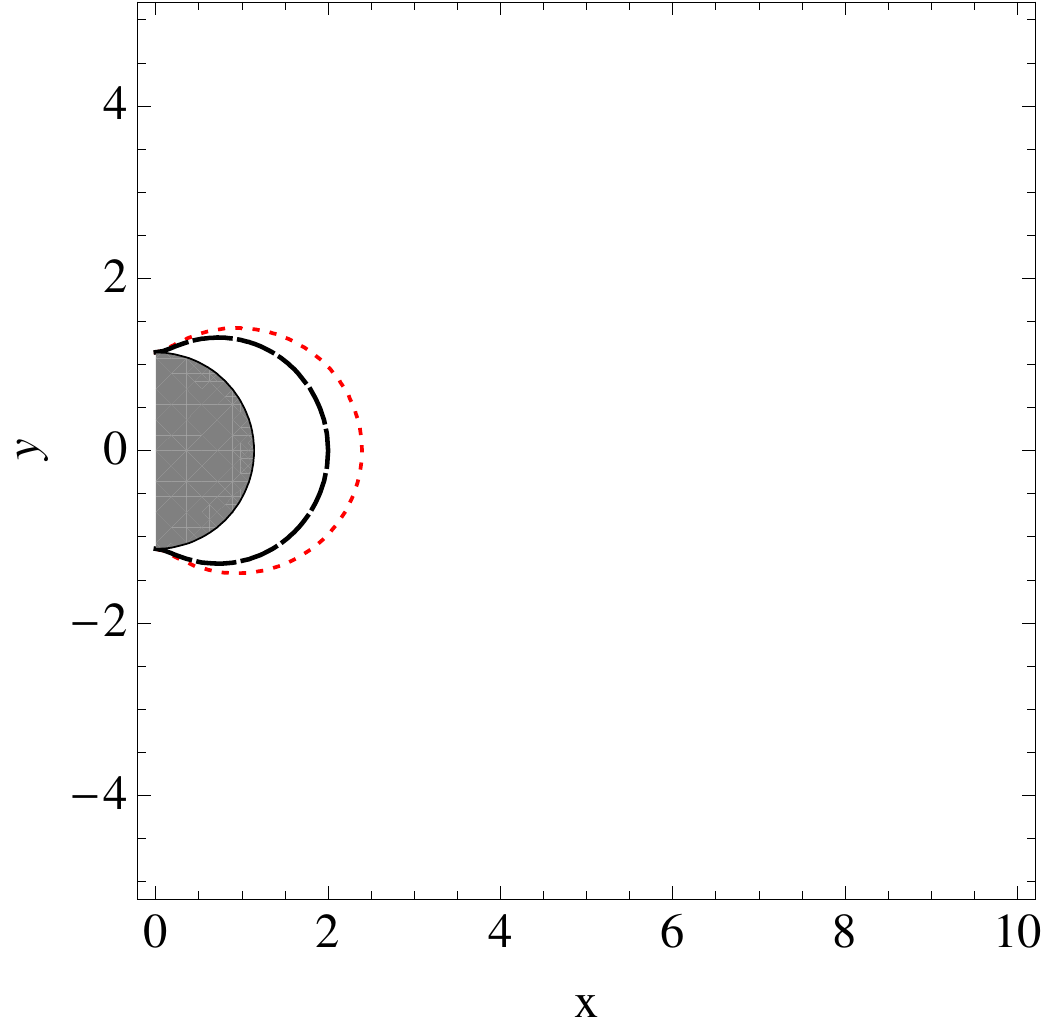}}
\caption{\label{Eoutfig2}  Magnified segments of Fig.~\ref{E50w0} and \ref{E50w1} which demonstrate the existence of yet a third boundary of motion for $\omega=0,-1$. This acts as a second outer boundary, therefore forbidding any string starting from inside the region bounded by the innermost red dotted curve to ever leave this region.}
\end{figure}

We now demonstrate how one can infer all the qualitative information about the boundaries of the motion from Fig.~\ref{Ejsq}.
For a given energy or value of $J$, there is a corresponding point on the solid black curve.  
This point can lie in several distinct regions of the graph, which correspond to qualitatively different types of
boundary:
\begin{enumerate}[(i)]
\item\label{item1} A region outside the shaded area and above the dashed red line. In this case there is 
just an outer boundary, {\em i.e.}~there is only one value of $r$ outside the horizon that yields a 
solution to eq.~(\ref{VKerr0}) for each $\theta$ for given $E$ and $J$.  The string  can potentially escape in the $y$ direction, as its energy is higher than the $E_{\rm esc}$.
\item A region outside the shaded area and below the dashed red line. Again, there is 
just an outer boundary. 
However, now the string  cannot escape in the $y$ direction as its energy is lower than the $E_{\rm esc}$.
\item A region inside the shaded area and above the dashed red line. In this case 
there are always outer and inner boundaries, 
and for the examples with $\o=0,-1$ there is 
yet another outer boundary at a lower value of $x$. 
That is,
eq.~(\ref{VKerr0}) has up to three solutions 
for $r$ outside the horizon
 each value of $\theta$ and for given $E$ and $J$. However, there is no boundary on the 
$y$ motion
and the string can potentially escape.
\item  A region inside the shaded area and below the dashed red line. 
The boundaries of the motion are the same in number as 
in the previous case, but 
with a qualitative difference: the inner and the outer boundaries at higher 
values of $x$ meet at a certain value of $y$ and together form a closed curve. 
Therefore, there is also effectively a boundary in the $y$ motion and the string 
can be trapped  in some toroidal volume. 
\end{enumerate}
There is a subtle point that should not pass unnoticed: 
depending on the initial conditions, 
a string with given $E$ and $J$ 
which corresponds to a point in the shaded region in Fig.~\ref{Ejsq} can either be moving 
in between the inner and outer boundary at higher $x$'s, or it can be 
inside 
the second outer boundary at smaller $x$'s. 
Finally it should be stressed that a second outer boundary at small $x$'s can 
only exist when there $E_b$ has a maximum outside the horizon.

\begin{figure}[t ]
\subfigure[~$E=38$, $\omega=1$ \label{E38wm1}]{\includegraphics[width=4cm]{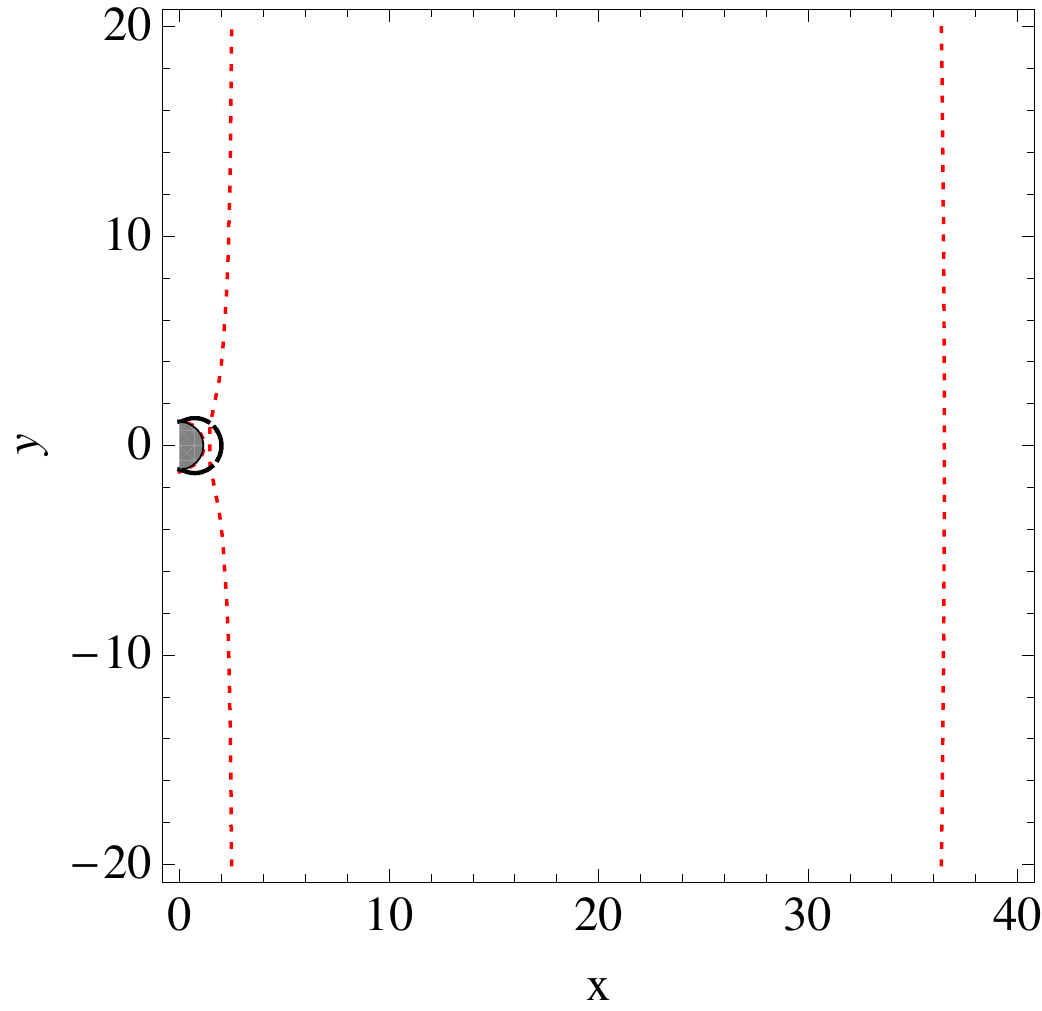}}
\subfigure[~$E=8$, $\omega=1$ \label{E8wm1}]{\includegraphics[width=4cm]{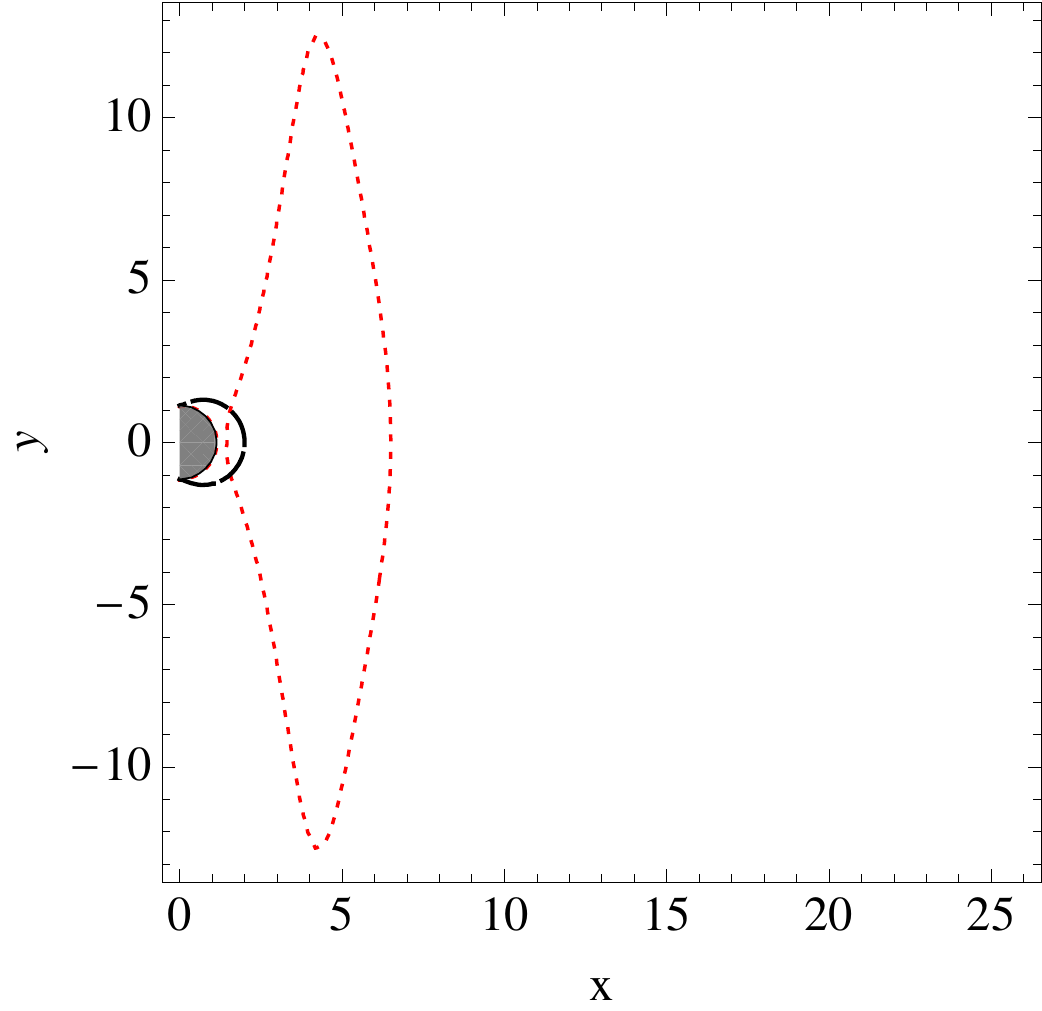}}\\
\subfigure[~$E=38$, $\omega=1$ \label{E38wm1z}]{\includegraphics[width=4cm]{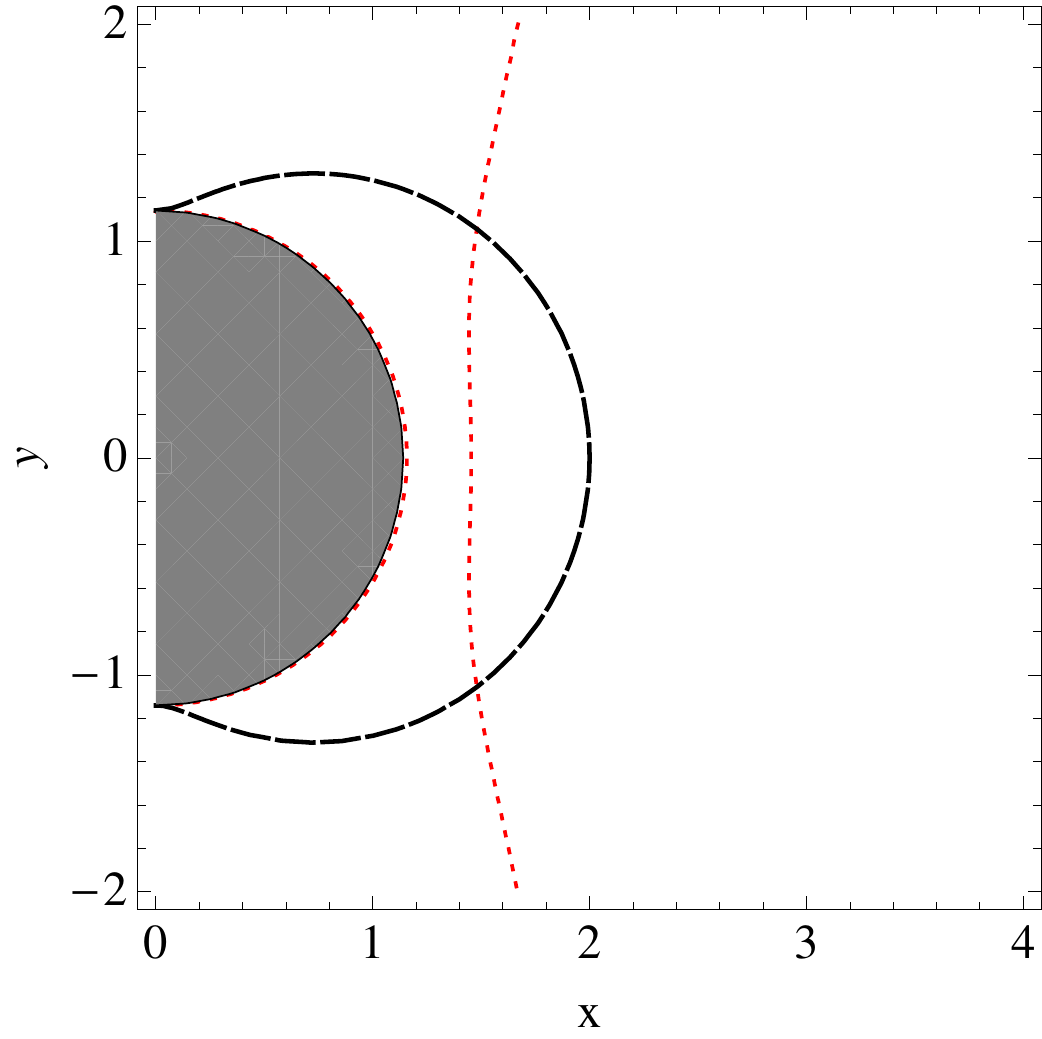}}
\subfigure[~$E=8$, $\omega=1$ \label{E8wm1z}]{\includegraphics[width=4cm]{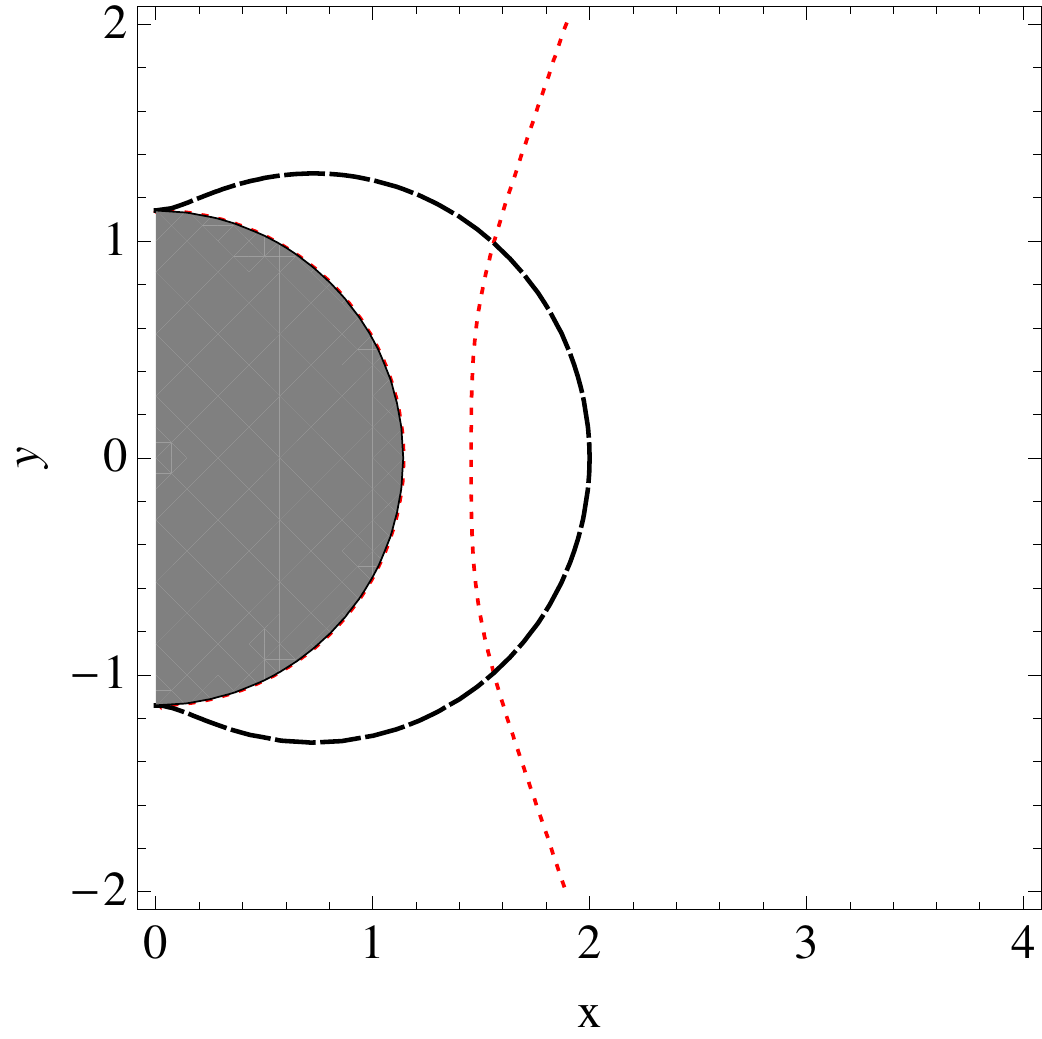}}\\
\subfigure[~$E=38$, $\omega=0$ \label{E38w0}]{\includegraphics[width=4cm]{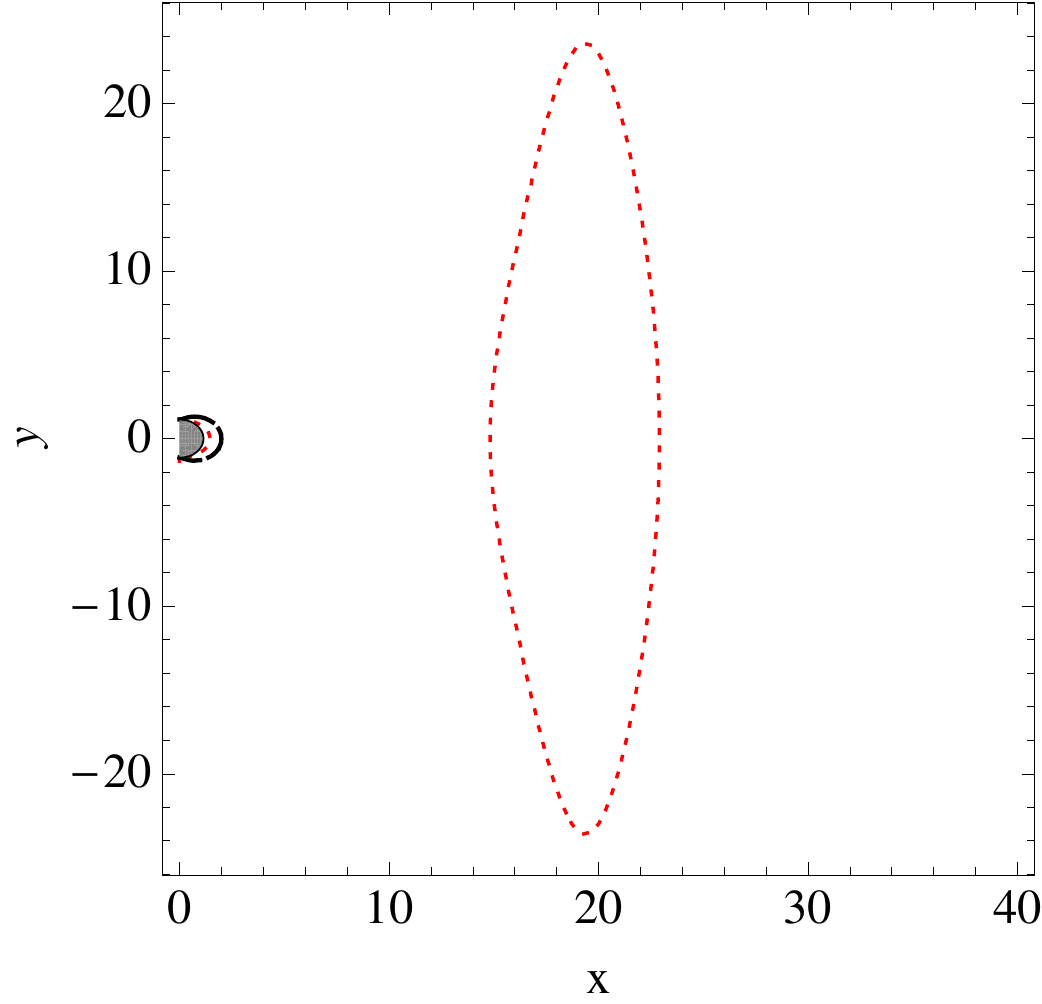}}
\subfigure[~$E=8$, $\omega=0$ \label{E8w0}]{\includegraphics[width=4cm]{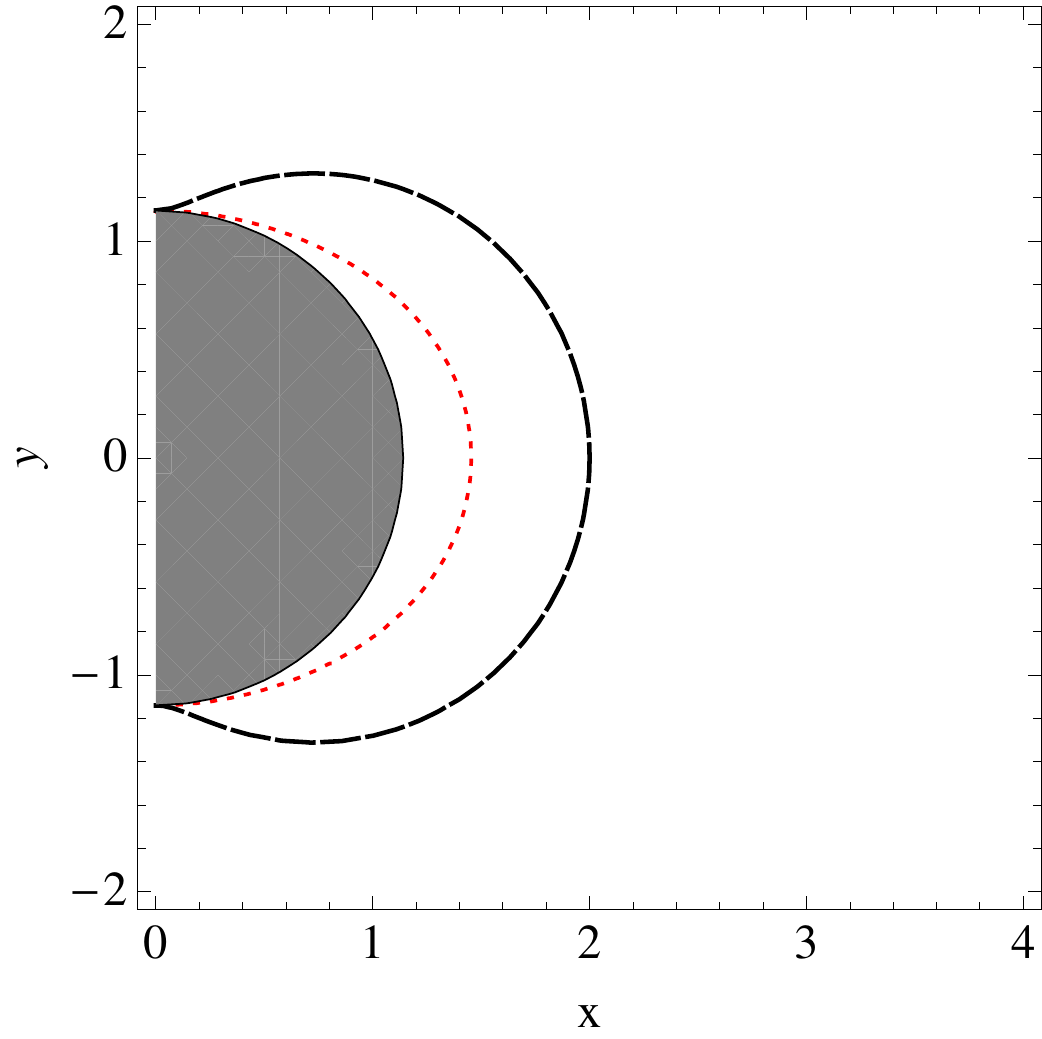}}
\caption{\label{Einfig}   Boundaries of motion for different values of the energy and the $\omega$ parameter, for  string starting at rest at $r=r_{\rm ISCO}$ and $\theta=\pi/2+0.02$.  The form of the boundaries is exactly what one could have predicted by using Figs.~\ref{Ejqswm1in}-\ref{Ejqsw1in}. Figs.~\ref{E38wm1z} and \ref{E8wm1z} focus on the region close to the black hole of Figs.~\ref{E38wm1} and \ref{E8wm1} respectively. Clearly, the inner boundary of motion lies within the ergoregion.}
\end{figure}

Next we give in Fig.~\ref{Eoutfig} a few examples of plots of the boundaries of motion corresponding to 
given initial conditions and current (or energy), for $r=r_h+1.25$. 
The qualitative nature of these boundary plots can 
be read off from Fig.~\ref{Ejsq}.
For example, with $E=50$ there exist both inner and outer boundaries in the $x$ direction 
for all values of $\o$, corresponding to the fact that the solid black curve
in Figs.~\ref{Ejqswm1}, \ref{Ejqsw0}, \ref{Ejqsw1} falls inside the shaded region when $E=50$. 
Additionally, for $\omega=1,0$ there is no boundary along the $y$ direction, 
whereas for $\omega=-1$ there is such a boundary, corresponding to the fact that
in the latter case the $E=50$ point on the
solid black curve lies below the dashed red line in Fig.~\ref{Ejqsw1}. 
In Fig.~\ref{Eoutfig2}, one can see the region closer to the black hole: for the $\omega=0,-1$ 
cases, there exists a third boundary, which is an outer boundary. This is in accordance with 
the form of $E_b$ in Figs. \ref{Ebw0} and \ref{Ebw1} 
and highlights the subtlety discussed above right after outlining the various regions of Fig.~\ref{Ejsq}.
 For $E=17$ on the other hand, only the $\omega=1$ choice 
 allows motion which is unbounded in the $y$ direction. For $\omega=0$ 
 the string is trapped in a toroidal volume, and for $\omega=-1$ there is 
 only an outer boundary. 
Fig.~\ref{Einfig} is similar to Fig.~\ref{Eoutfig}, but for $r=r_{\rm ISCO}$. (We do not plot the $\o=-1$ 
case since
there is no energy for which there will be an internal boundary for the motion.) 
In this case,
the inner boundary of motion lies within the ergoregion. 

Before closing this section it is worth stressing once more the similarity between 
the contour plots of the boundaries of motion for a Newtonian elastic spinning ring 
presented in Fig.~\ref{fig4} and the boundaries of motion for the relativistic strings presented in this section.

\subsubsection{Trajectories}

In this subsection we look at examples of individual trajectories, and in particular 
we exhibit processes whereby a string is ejected along the axis, with some of the
initial internal energy of the string converted to translational kinetic energy.
To this end, we employed numerical integration with Mathematica 6.0
to solve the equations of motion.
As already mentioned we consider here two different initial conditions:
one for which the string starts outside the ergoregion, and one 
for which the string starts at $r=r_{\rm ISCO}$, which lies well within the 
ergoregion (points $P_1$ and $P_2$ in Fig.~\ref{figinit} respectively).
Starting at rest from these initial positions, 
there are 
three types of trajectories, as 
indicted by the types of 
boundaries of motion: 
those that 
are trapped inside an outer boundary with no
inner boundary and simply
fall in across the horizon, 
those that are trapped in some toroidal volume, 
and those that can reach spatial infinity
along the direction of the axis. The trajectories of the first type
display no interesting features so we do not present any 
explicit examples of them 
here.\footnote{Note that one can have trajectories of this type 
even if there exists an inner boundary, provided that an outer boundary 
at smaller $x$'s exists as well, as discussed earlier. This is the case, 
for instance, when the boundaries of motion are those of 
Fig.~\ref{E50w0zoom} or Fig.~\ref{E17w0} and the string start at 
rest at $r=r_h+1.25$.}

%

\begin{figure}[t]
\subfigure[~$E=9$, $r=r_h+1.25$ \label{E9wm1mot}]{\includegraphics[width=4cm]{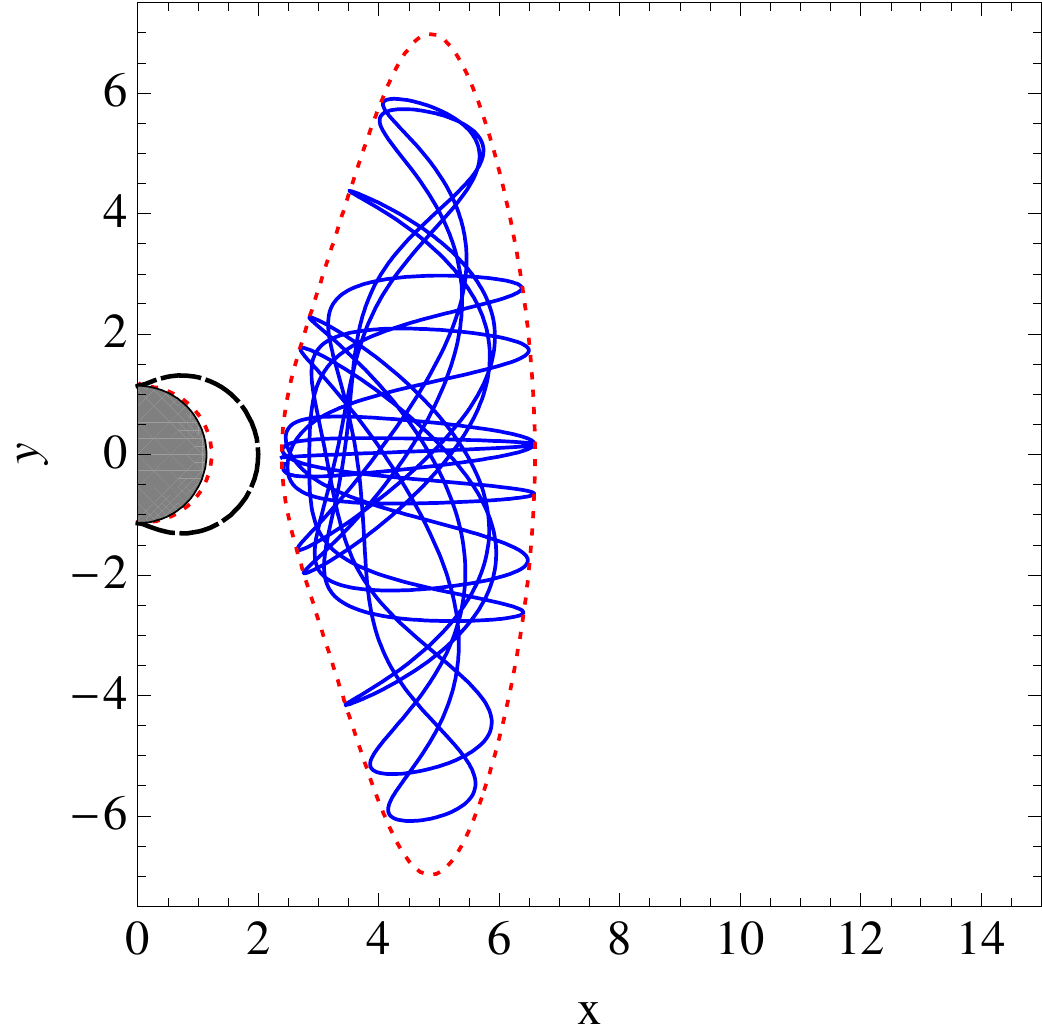}}
\subfigure[~$E=9$, $r=r_h+1.25$  \label{E9wm1motz}]{\includegraphics[width=4cm]{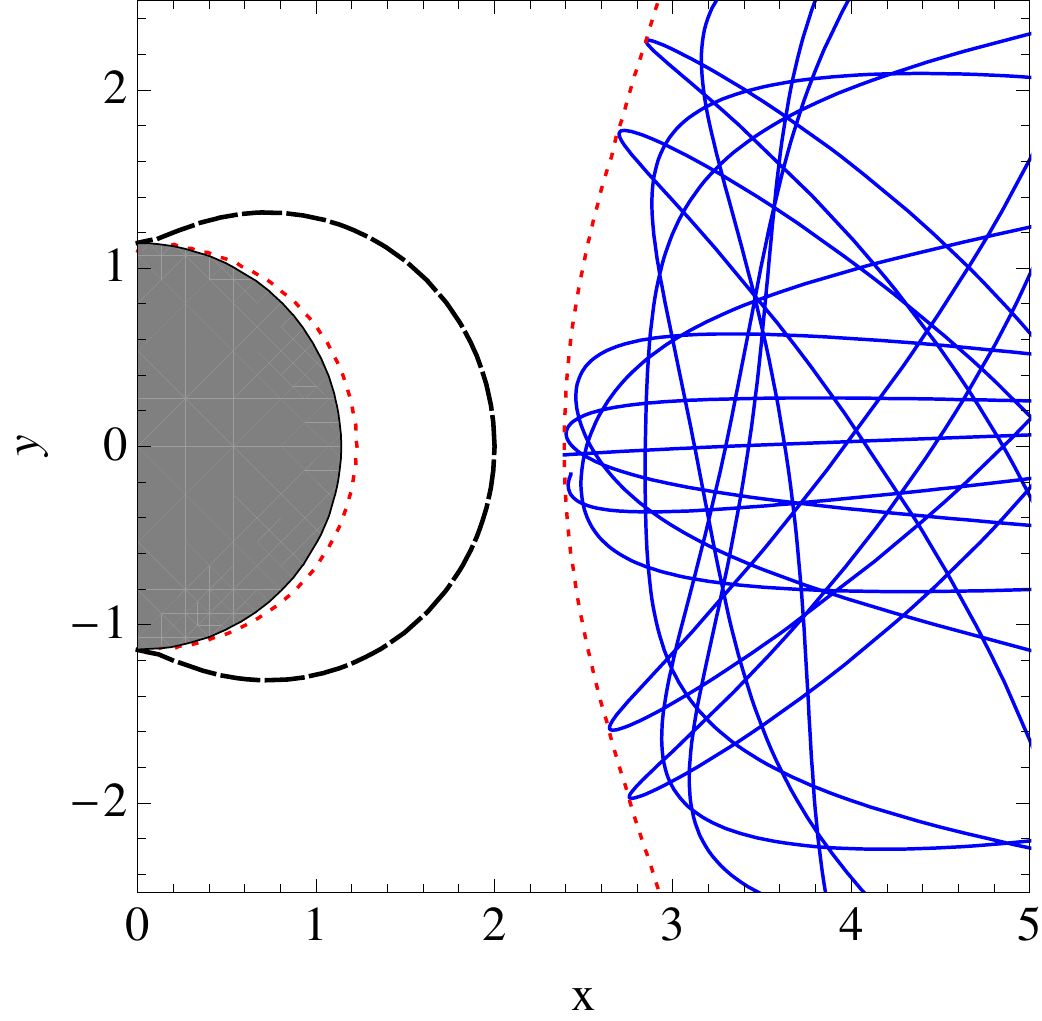}}\\
\subfigure[~$E=8$, $r=r_{\rm ISCO}$  \label{E8wm1inmot}]{\includegraphics[width=4cm]{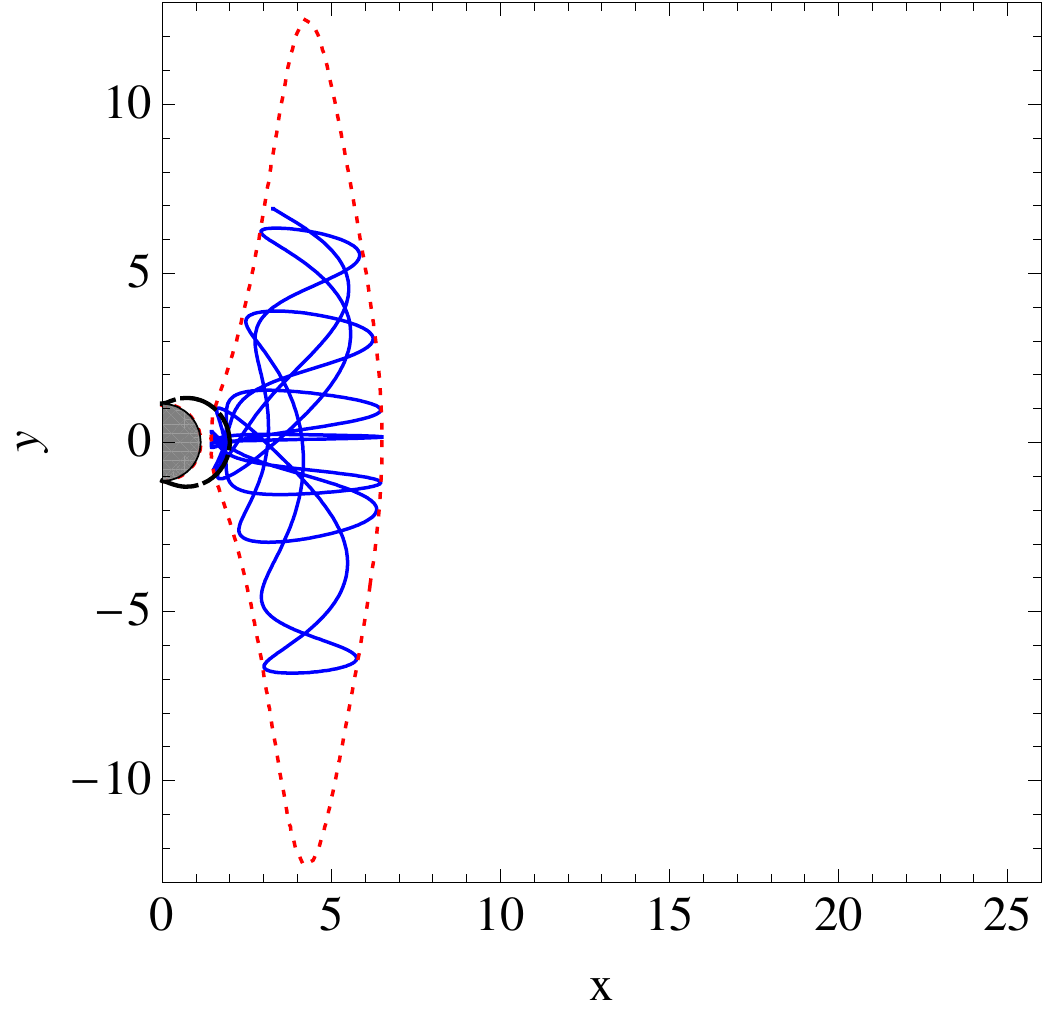}}
\subfigure[~$E=8$, $r=r_{\rm ISCO}$  \label{E8wm1inmotz}]{\includegraphics[width=4cm]{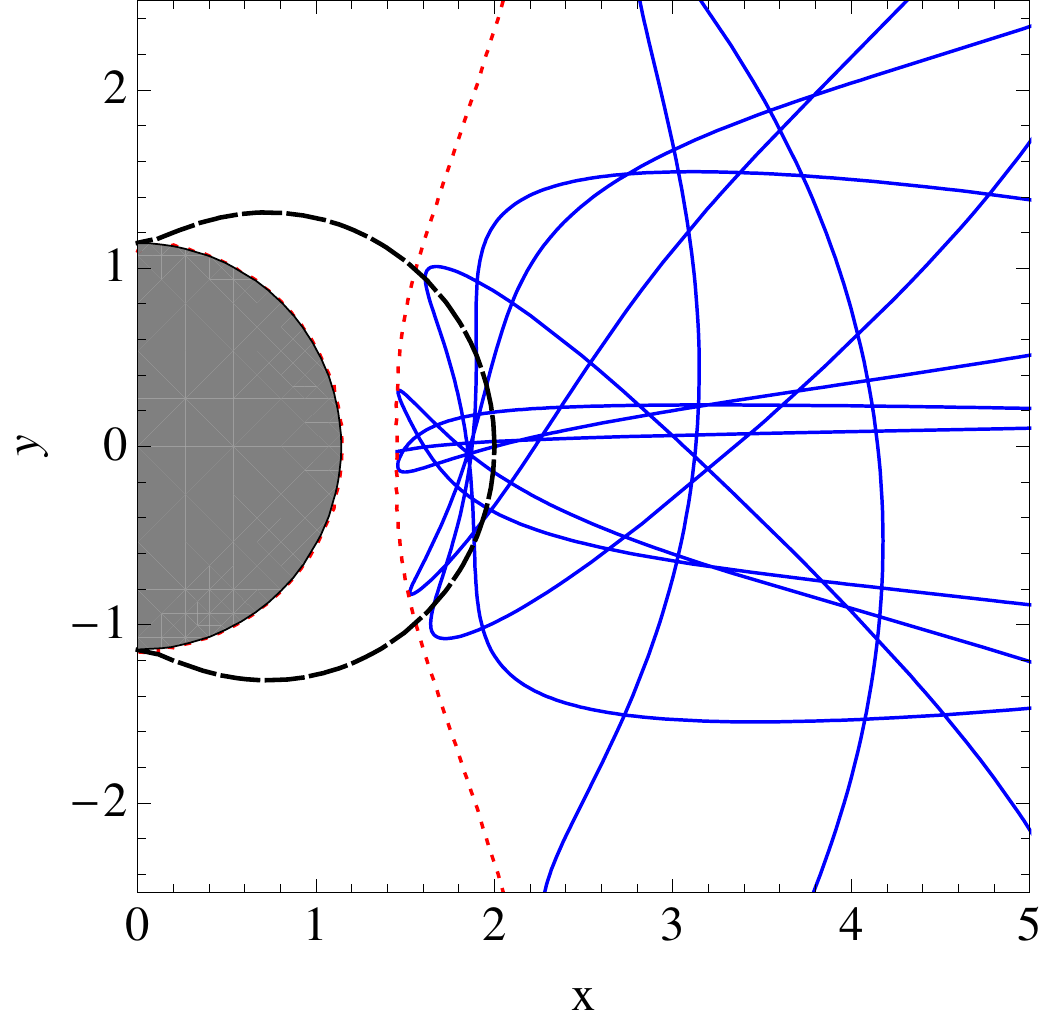}}
\caption{\label{boundmotionfig} Confined trajectories of the string for different initial positions and different energies. 
The blue continuous line corresponds to the trajectory of a point on the string ($\phi$ dimension suppressed). 
No qualitative difference seems to exist between the motion of a string starting inside the ergoregion and 
one starting outside. In both case the motion seems to follow no clear pattern.}
\end{figure}

\begin{figure}[t ]
\subfigure[~$E=8$, $r=r_h+1.25$ \label{E8wm1mot}]{\includegraphics[width=4cm]{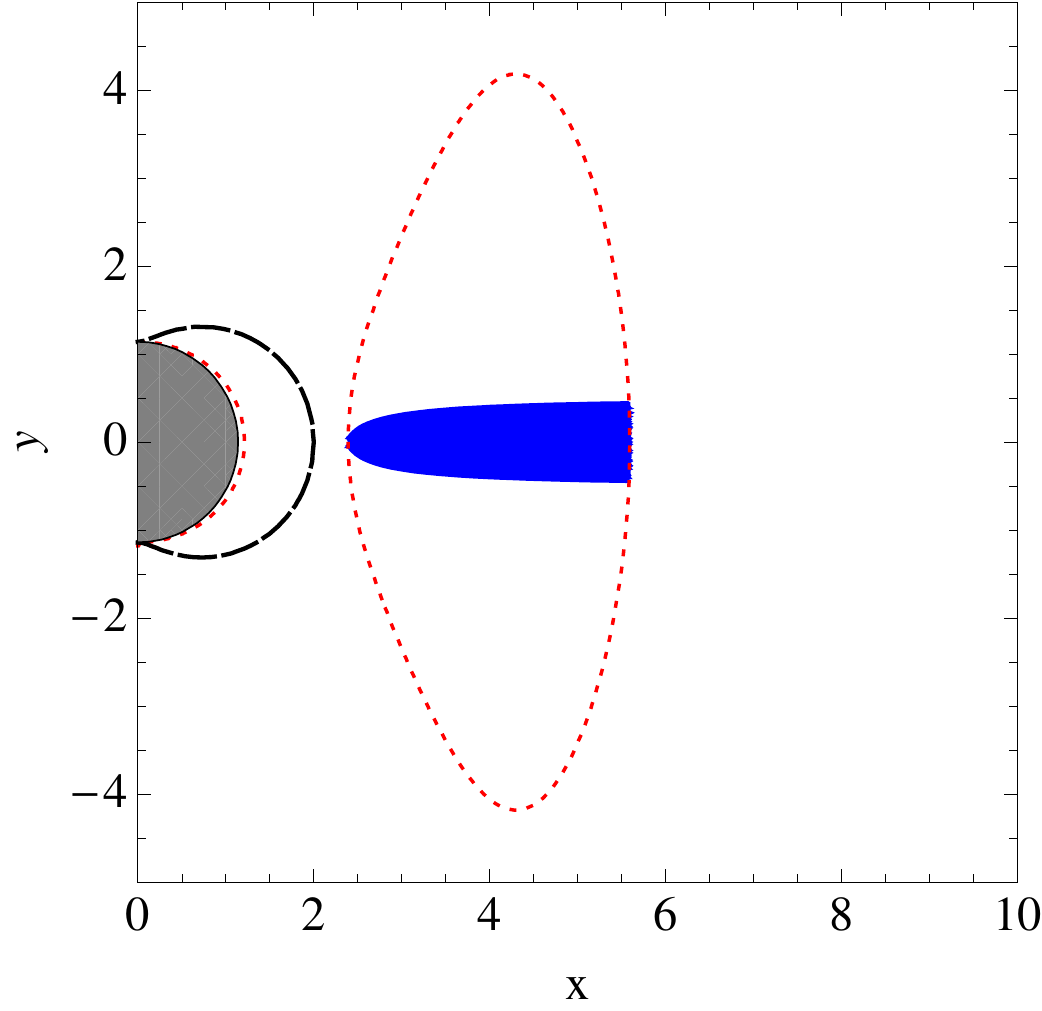}}
\subfigure[~$E=8$, $r=r_h+1.25$  \label{E8wm1motz}]{\includegraphics[width=4cm]{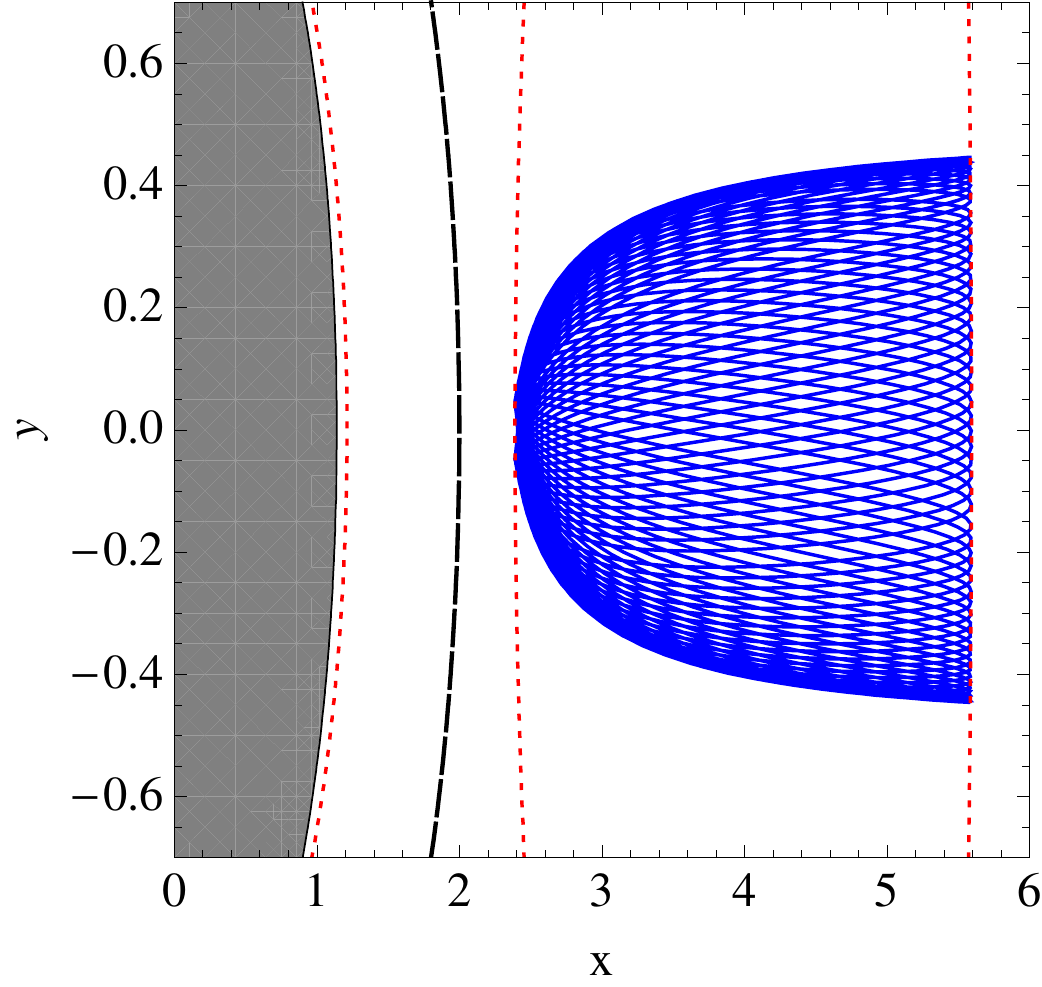}}\\
\subfigure[~$E=5.5$, $r=r_{\rm ISCO}$  \label{E5.5wm1inmot}]{\includegraphics[width=4cm]{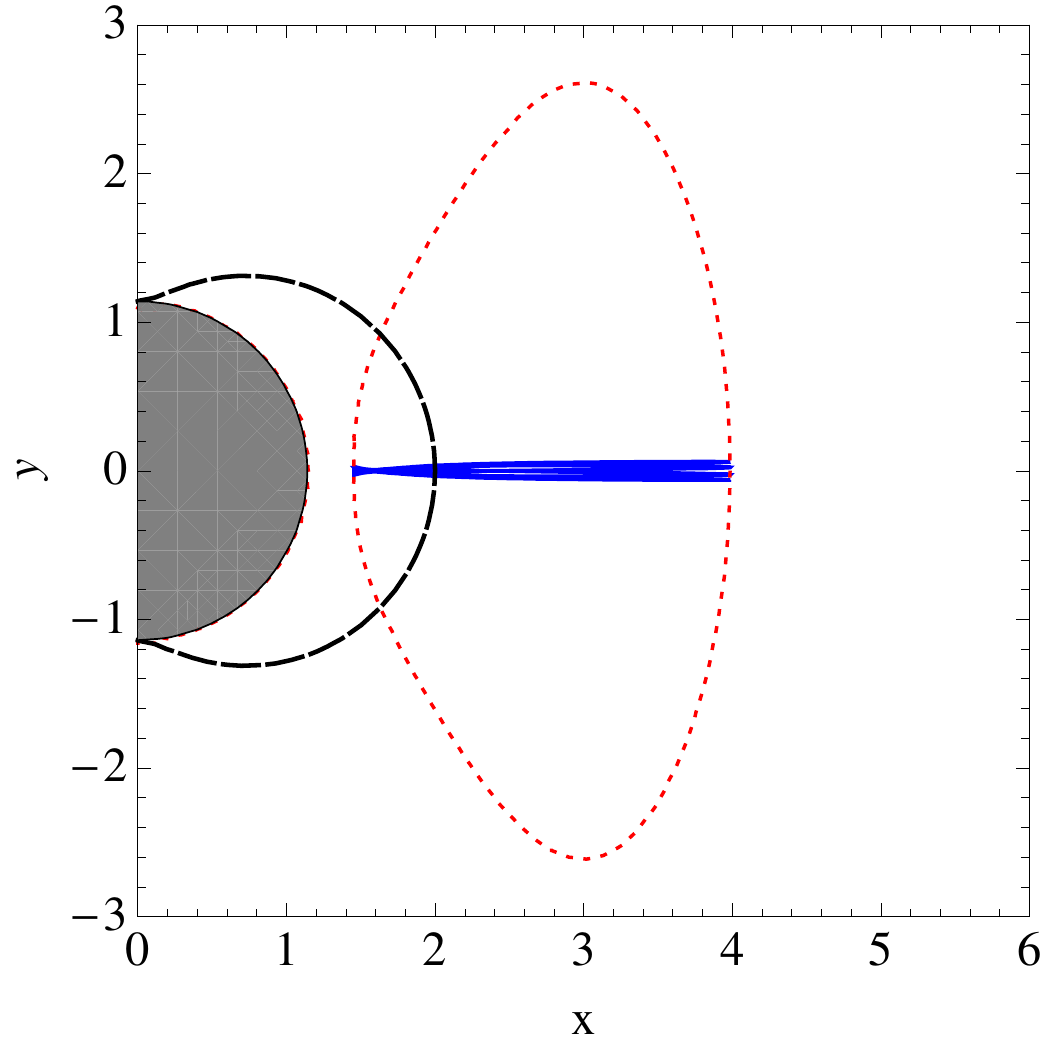}}
\subfigure[~$E=5.5$, $r=r_{\rm ISCO}$  \label{E5.5wm1inmotz}]{\includegraphics[width=4cm]{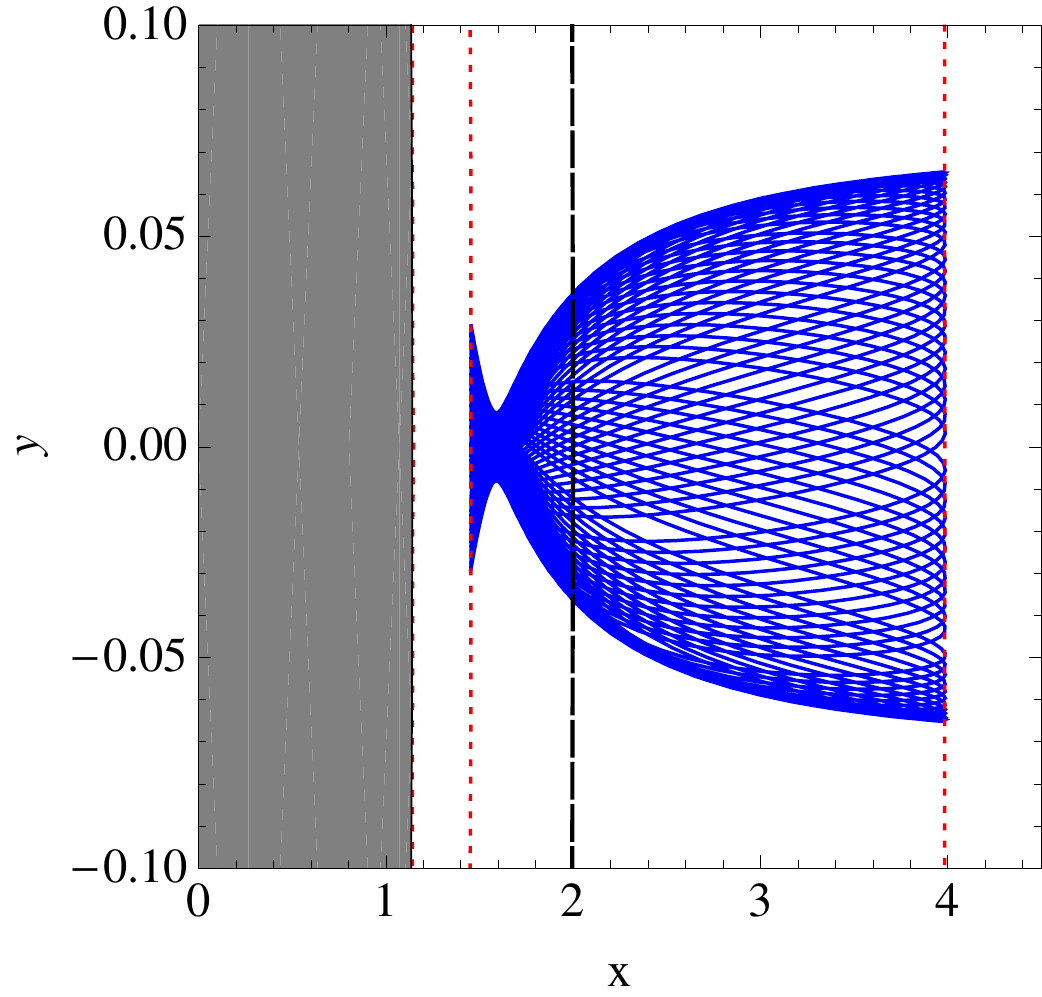}}
\caption{\label{focusfig} ``Focusing'' of the trajectories. In this case a pattern seems to be arising. In Figs.~\ref{E8wm1motz} and \ref{E5.5wm1inmotz} we have distorted the aspect ratio in order to make this manifest.}
\end{figure}

We begin by discussing 
bound trajectories. 
The energies we considered previously 
do not
lead to confined motion for a string starting at $P_1$. 
This can be seen in Fig.~\ref{Eoutfig}: For the cases where there can be confined motion, 
namely Figs.~\ref{E17w0} and \ref{E50w1}, our initial position lies on the innermost 
outer boundary. Thus the string would fall straight into the black hole. 
However, for other energies there can be 
confined motion for a string starting at $P_1$. Also, according to Fig.~\ref{E8wm1} 
we have already found an example where the string's motion is confined when it starts 
from $P_2$, {\em i.e.}~inside the ergoregion. In Fig.~\ref{boundmotionfig} we present 
the motion of a string starting from $P_1$ with $E=9$ and that of a string starting from 
$P_2$ with $E=8$ for qualitative comparison. Both cases refer to $\omega=1$ (co-rotation). 
In both cases the strings move somewhat randomly, without following 
any clear pattern. 
Fig.~\ref{E9wm1motz} and \ref{E8wm1inmotz} focus on the area inside and around the 
ergoregion. 
The string is sometimes pulled by gravity rather 
sharply back toward the equatorial plane after reflecting from the inner boundary.
We have monitored the accuracy by running the code backward in time from the point the 
forward run ended, and checking how closely this backward run 
approaches 
the initial conditions. 
In the plots presented we stop the simulation before the accuracy becomes lower than 
$10^{-4}$.\footnote{Using only the default settings of Mathematica, we
were able to run the simulation for $\tau\sim 150-200$ for trajectories starting from 
$P_1$ and for $\tau\sim 50-100$ for trajectories starting from $P_2$ (provided that 
the string did not end up in the inside the horizon before that time) before dropping 
below the required accuracy. Remarkably, for the ``focused" trajectories, such as 
those in Fig.~\ref{focusfig}, the accuracy was still two orders of magnitude higher 
than our minimum requirement at $\tau=500$. Note that $\tau$ is the coordinate 
time on the worldsheet.}

Before proceeding further to trajectories that can reach spatial infinity, 
there is another interesting feature of confined motion which should not be missed. 
In Fig.~\ref{focusfig} we present confined trajectories starting from $P_1$ with 
$E=8$ and from $P_2$ with $E=5.5$, always for $\omega=1$. There 
is evidently
some remarkable ``focusing" of the motion close to the equatorial plane in both cases,
and a very regular pattern is generated. In 
Figs.~\ref{E8wm1motz}
and \ref{E5.5wm1inmotz} we have distorted the aspect ratio 
in order to make manifest the pattern that arises in the motion. 
We do not understand why these cases are so different from those shown in Fig.~\ref{boundmotionfig}. The focussing cannot be explained by considering only the boundary of allowed motion (dotted curve) as inferred using the effective potential.  The appearance of a Lissajous-like pattern indicates that when the string stays  close to the equatorial plane the motion becomes similar to complex periodic oscillation.

\begin{figure}[t ]
\subfigure[~$E=17$, $r=r_h+1.25$ \label{E17wm1mot}]{\includegraphics[width=4cm]{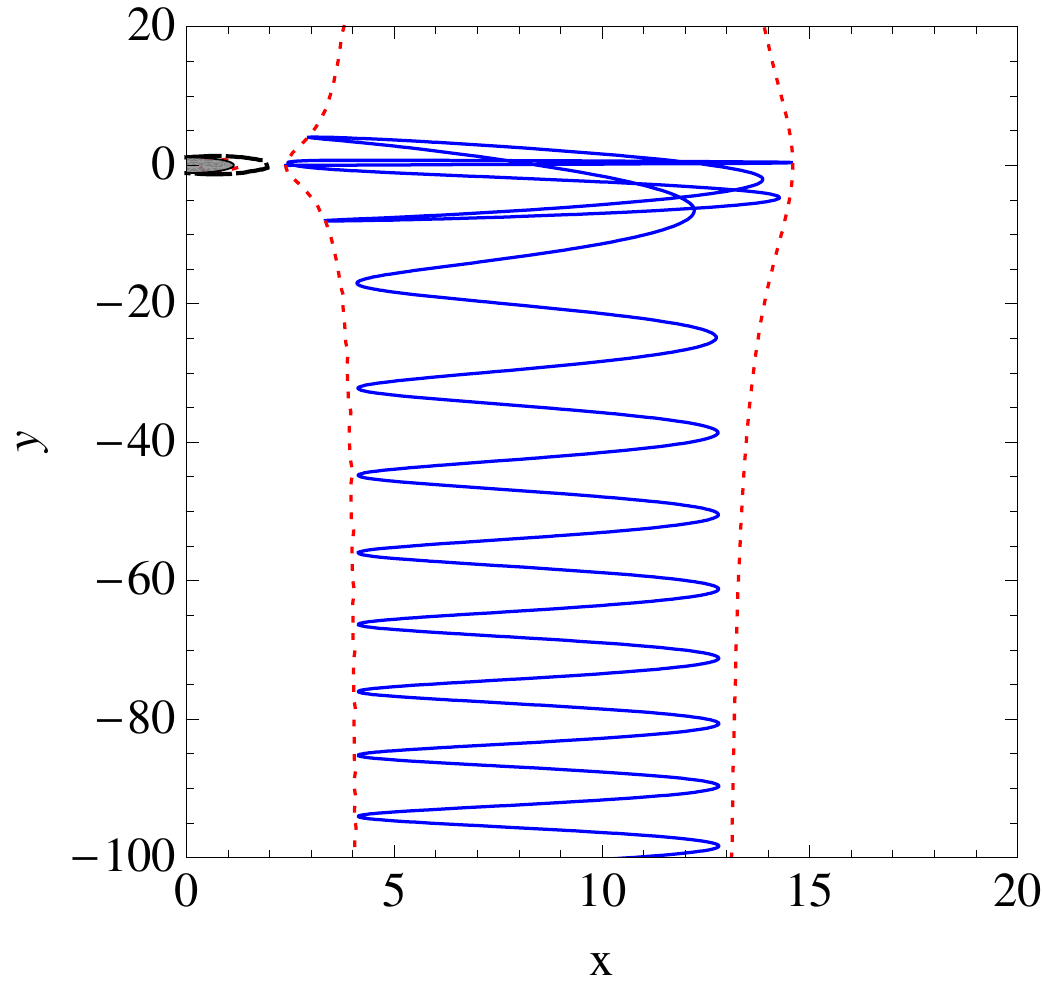}}
\subfigure[~$E=17$, $r=r_h+1.25$  \label{E17wm1motz}]{\includegraphics[width=4cm]{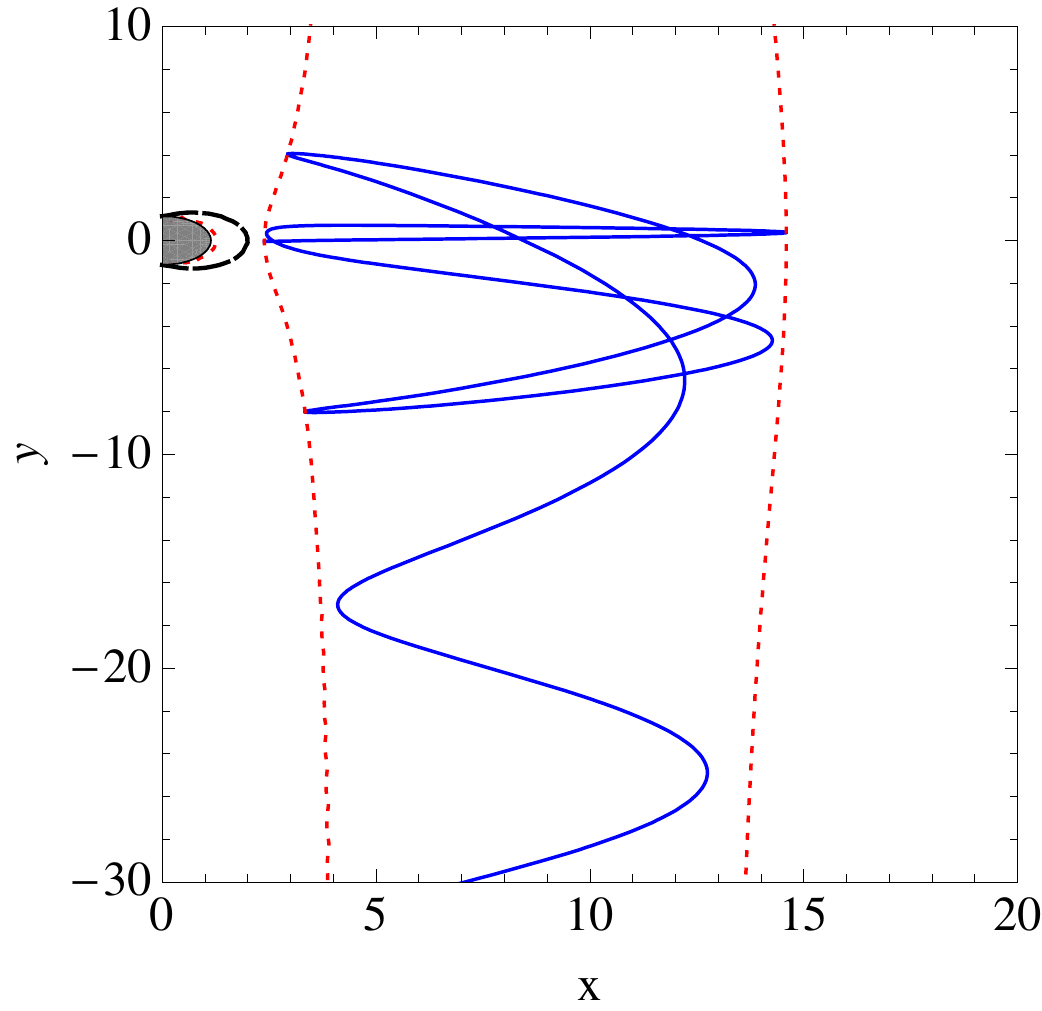}}\\
\subfigure[~$E=38$, $r=r_{\rm ISCO}$  \label{E38wm1inmot}]{\includegraphics[width=4cm]{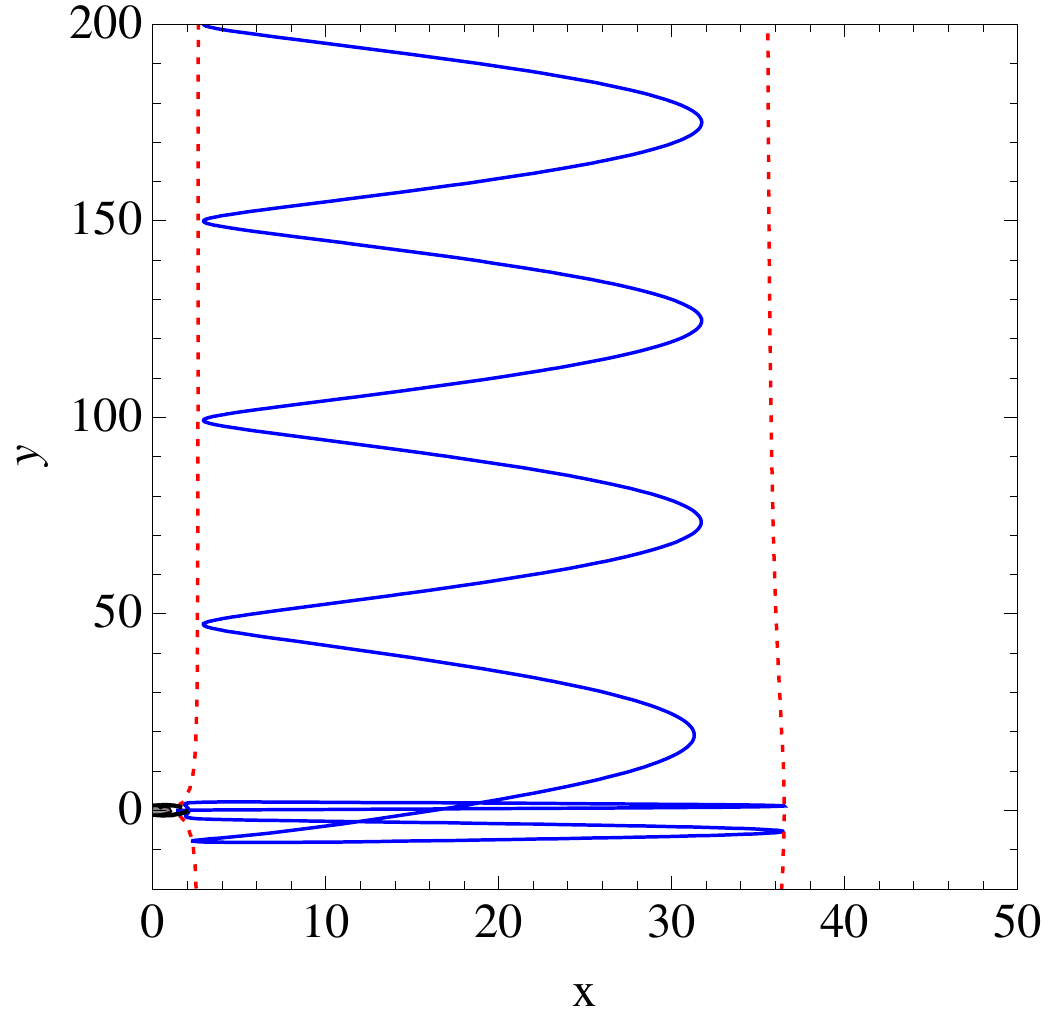}}
\subfigure[~$E=38$, $r=r_{\rm ISCO}$  \label{E38wm1inmotz}]{\includegraphics[width=4cm]{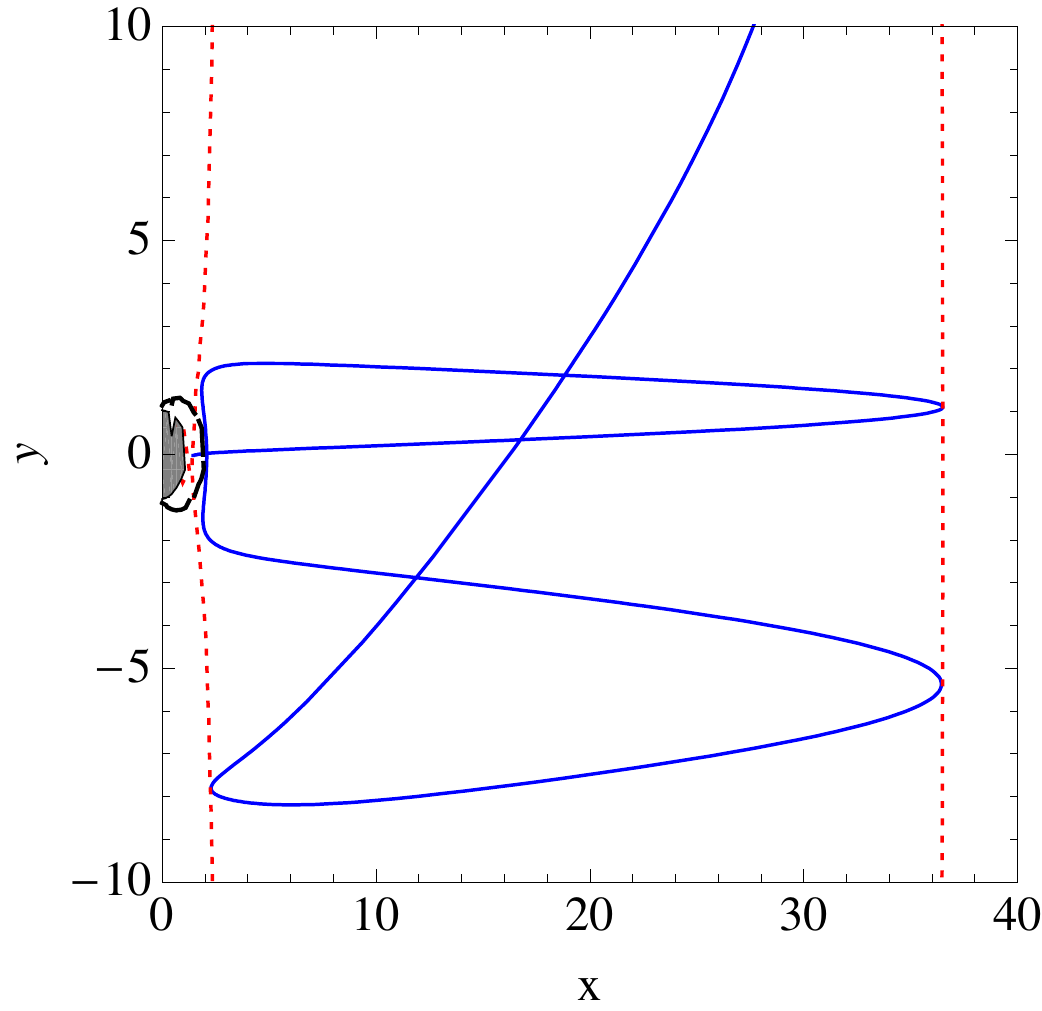}}
\caption{\label{freefig} Escape trajectories for strings of different energy and angular momentum, 
start at $P_1$ and $P_2$ respectively. After some oscilations around the equatorial plane the 
strings escape to infinity along the axis of symmetry. Clearly there is some conversion of 
internal to kinetic energy.}
\end{figure}

Finally, 
we discuss
trajectories that can reach spatial 
infinity.\footnote{These trajectories are similar to the 
time-reversed version of Larsen's adiabatic capture trajectories found in Ref.~\cite{Larsen:1993nt}.}
In Fig.~\ref{E17wm1mot} we have plotted the trajectory of a string starting from $P_1$ with $E=17$ 
and $\omega=1$. After a couple of oscillations around the equatorial plane, which are exhibited 
more clearly in Fig.~\ref{E17wm1motz}, the string escapes along the $z$ direction. 
The behavior of a string starting from $P_2$ with $E=38$ and $\omega=1$ is quite similar. 
In both cases, there is some conversion of internal energy of the string into kinetic energy along the $y$ axis.

\begin{figure}[t ]
\subfigure[~$E=11.3$, scattering \label{scattering}]{\includegraphics[width=4cm]{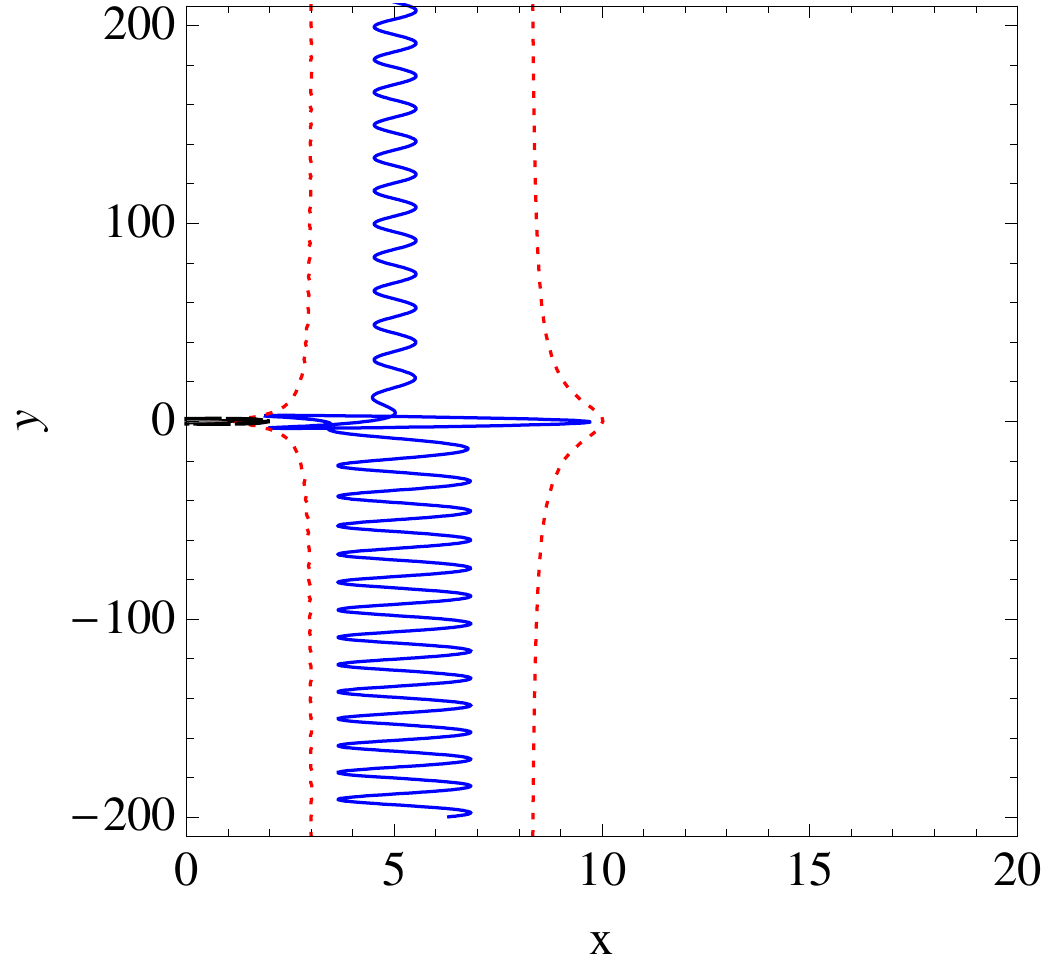}}
\subfigure[~$E=11.3$, scattering  \label{scattzoom}]{\includegraphics[width=4cm]{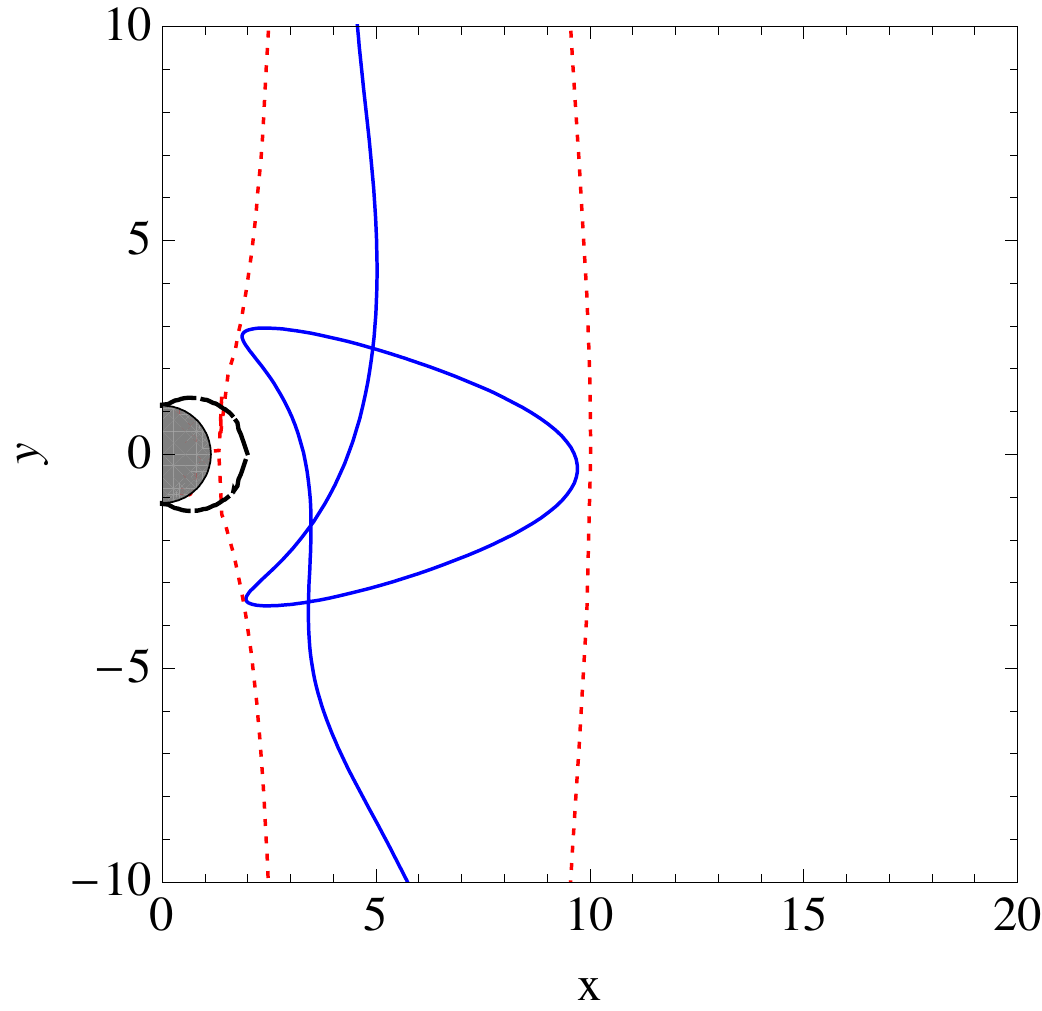}}\\
\subfigure[~$E=11.8$, collapse  \label{collapse}]{\includegraphics[width=4cm]{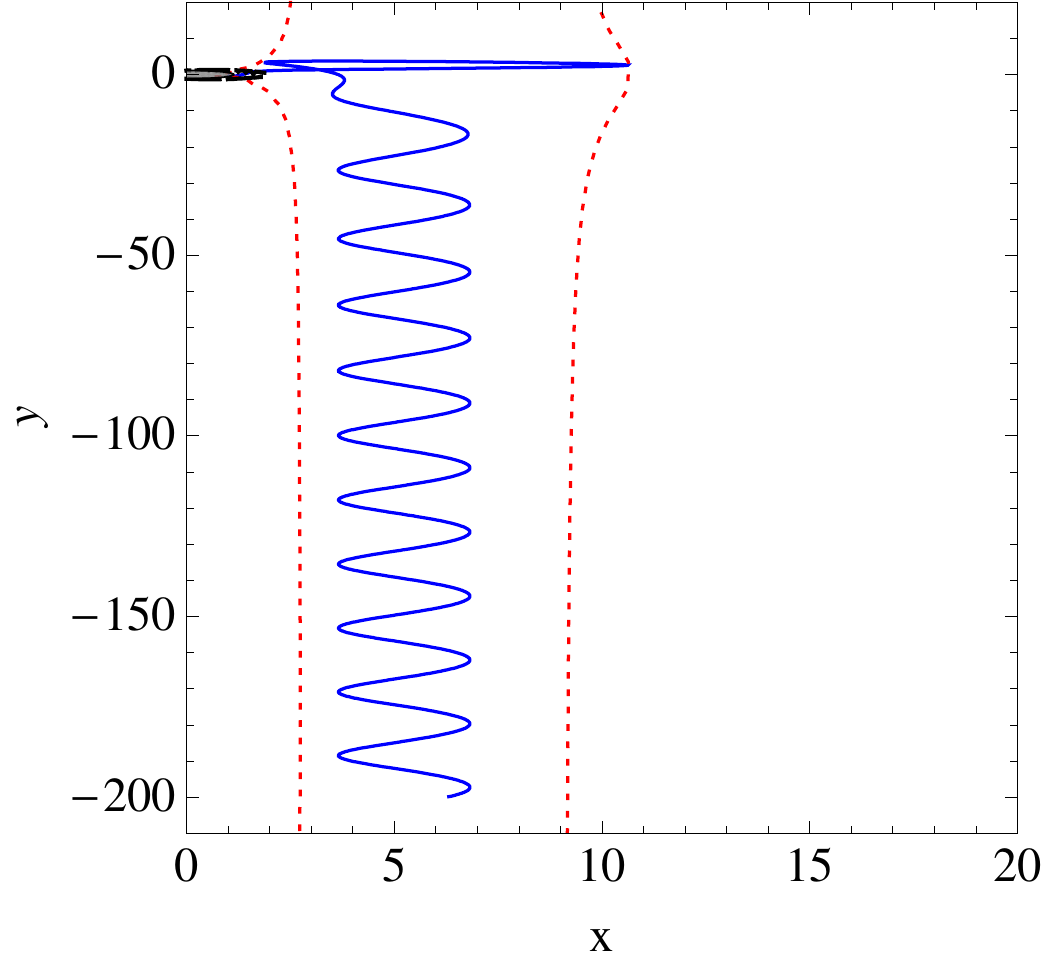}}
\subfigure[~$E=11.8$, collapse  \label{collapsezoom}]{\includegraphics[width=4cm]{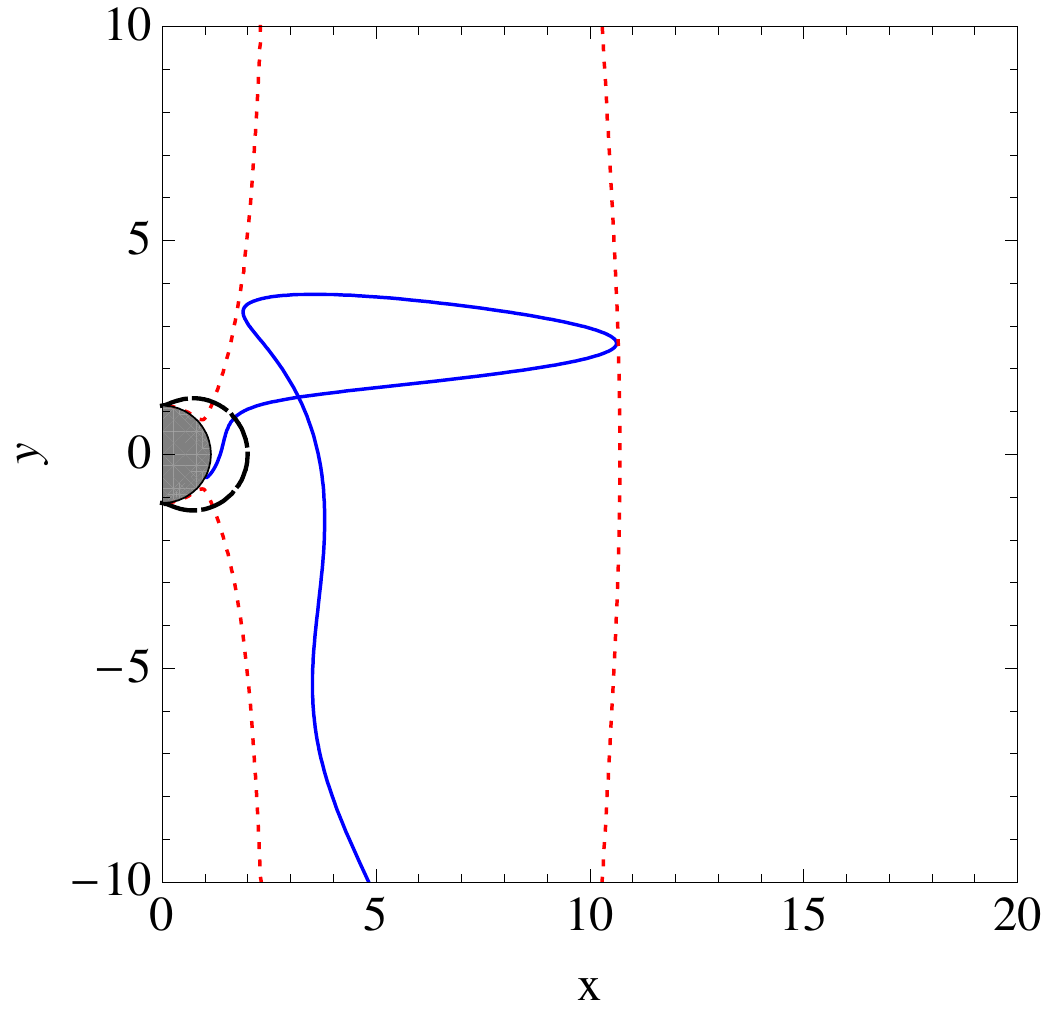}}
\caption{\label{scattcaptfig} Scattering and collapse on the black hole of strings coming from spatial infinity. 
The scattering can lead to energy conversion, as can 
be seen from Fig.~\ref{scattering} by the fact that the 
average radius of oscillation along the $x$ direction has been
slightly reduced after the scattering.
 On the other hand, if the inner boundary of motion lies within the horizon close to 
 the equatorial plane, then instead of a scattering there can be a collapse, see 
 Figs.~\ref{collapse} and \ref{collapsezoom}. }
\end{figure}

Another interesting class of non-confined trajectories are 
the scattering trajectories
found by Larsen in Ref.~\cite{Larsen:1993nt}. 
An example of such a trajectory for a string with $E= 11.2929$, $\omega=1$, and initial conditions  $r_0=200$, $\theta_0=3.11$, $\dot{r}_0= -3.4$ and $\dot{\theta}_0=-0.007$ is shown in Fig.~\ref{scattering}. 
The string comes in along the axis and after a couple of bounces
around the black hole, 
which can be better appreciated in Fig.~\ref{scattzoom}, escapes
to infinity out the other side.
The scattering 
in this case reduces the mean radius in the 
$x$ direction, signaling 
a conversion of 
internal
energy to kinetic energy along the $y$ direction,
as explained below in the next subsection, \ref{ejection}.

In Figs.~\ref{collapse} and \ref{collapsezoom} we show a case where the string 
actually falls into the black hole instead if getting scattered. For this case, 
$E= 11.8417 $, $\omega=1$, and the initial conditions  are $r_0=200$, $\theta_0=3.11$, 
$\dot{r}_0= -2.4$ and $\dot{\theta}_0=-0.007$. In this case 
the string has more energy than a string at rest on the horizon 
and, hence, 
its motion has no inner boundary close to the equatorial plane 
({\em cf.}~the discussion in section \ref{boundaries} about Figs.~\ref{Ebwm1} and \ref{Ejsq}).

\subsubsection{Ejection energy} 
\label{ejection}
As noted, on some trajectories internal energy of the string can be converted
to translational kinetic energy along the axis. To characterize this energy
we may associate a relativistic boost factor $\G=E/E_0$ with the asymptotic
string, where $E_0$ is the energy of the string in its center of mass frame
at infinity. In the center
of mass frame, all velocities vanish at the inner and outer boundaries 
$\r=\r_{i,o}$ of oscillation, so we may use (\ref{V0flat}) to obtain 
$E_0=\r  + J^2/\r$. Eliminating $J$ we find $E_0=\r_i+\r_o$. 
Thus, using the present notation $x$ for the cylindrical radius, we have  
\beq
\G=\frac{E}{x_i+x_o}.
\eeq
With this relation the $\G$ factors corresponding to the processes shown
in the previous examples can be computed. For example, for the $E=17$ case 
in Fig.~\ref{freefig} we find $\G\approx  1.01$, or $v/c\approx 0.15$, and 
for the $E=38$ case we find $\G\approx 1.09$, or $v/c\approx 0.39$.
We expect that these mildly relativistic boost factors
are typical of what one can expect in generic cases. 

\subsubsection{The role of black hole spin}\label{spin}

The black hole spin enters the string dynamics
via the $a$ dependence of the various metric
coefficients. While the value of $a$, certainly affects the boundaries of motion and dynamics when other quantities
are held fixed, we do not see any systematic qualitative effect
that can be attributed uniquely to the spin. That is,
by varying the current, energy, and location of the string similar 
effects can be produced. This statement is based both
on examining cases (not explicitly reported here) 
for which $a=0$, as well as by comparing the behaviors 
for different values of $\o$. The dynamics only depends
on $\o$ when $a$ is non-zero, but the effect of 
changing $\o$ can also 
generally be reproduced by changing other quantities.

The one exception to this statement concerns the
issue of negative Killing energy states. In the presence
of black hole spin, there is an ergoregion, inside of which
such states exist for point particles.
The same is true for the string, though the mechanism
is a bit different. 

For a particle, 
negative angular momentum orbits  in the ergoregion have negative 
Killing energy. 
Longitudinal motion of a string is meaningless, so it 
is not the ``orbital" motion, but rather the current 
on the worldsheet that can access negative energy
states. In particular,
the second term on the right hand side of 
the energy (\ref{VKerr0}), i.e.\ 
$-g_{t\phi}\tS^{\s\t}=|g_{t\phi}/g_{\phi\phi}| L/2\pi$,
is negative there when the angular momentum 
$L=4\pi\o J^2/(1+\o^2)$ is negative. 
The first term in eq.~(\ref{VKerr0}) vanishes
at the horizon, so the energy is certainly negative there
when $L<0$. 
The negative
energy states extend out 
a certain distance (which depends on the values 
of $\o$ and $J$) 
from the horizon into the ergoregion.
For example, one can see in Fig.~\ref{Ejqsw1in} 
the negative energy 
states at $r_{\rm ISCO}$ in the case $\o=-1$, except for the
smallest values of $J$.
(For the limiting case of zero tension and $\o=-1$, there
are negative energy states everywhere in the ergoregion.)

\section{Relevance to Astrophysics?}

As stated in the Introduction, 
although our primary aim
in this paper
was to study the motion of a current carrying strings in a Kerr background 
as an interesting dynamical system, 
our investigation was also loosely motivated 
by the problem of the production of collimated 
jets in astrophysics. 
We should 
therefore like to comment on what
might be the relevance of our findings to 
astrophysical systems. 

First of all, 
the current carried by the string 
in the model need not represent  
an electric current, but it can
just serve to model the 
role of 
angular momentum. In this sense, a current carrying string 
can serve as an idealization of some MHD matter configuration 
with tension and angular momentum, as mentioned in the Introduction. 
Simplistic as it may be,
this allows 
for the identification of a 
mechanism
that might conceivably be relevant:
the tension sets a barrier for the motion that keeps the string 
within a certain radius from the axis of symmetry, 
while
the angular momentum sets a barrier that keeps the string 
outside a certain radius.  Gravity deforms these 
barriers creating 
either a closed region in which the string is trapped, 
or a ``restricted path"  that leads to motion along the symmetry 
axis.
If some physical process were to 
``feed" such a 
system 
with strings close to the equatorial plane, then 
under the right circumstances one might end up with a very well 
collimated stream of strings moving out along the axis of symmetry.
Of course, the axisymmetry we are imposing might be too restrictive,
and in particular, even if a string started out approximately axisymmetric, 
it might wobble or twist and reconnect. Perhaps the tension 
could act to suppress such effects, but this requires further investigation.  
 
Still, one may speculate that the general flavor of the mechanism 
exhibited by the strings might generically be present in a 
class of astrophysical systems. 
For instance, given a system with an accretion disk, plasma
leaving the inner edge of the disk would carry some angular momentum,
and plasma flux tubes do possess a significant tension (which however 
is not constant but grows in proportion to the length).
It might even be that differential rotation can stretch such a tube,
storing in it a large amount of electromagnetic energy before launching
it towards the axis. External magnetic fields, which have been left
out of consideration in our initial study, might also play an important
role in the subsequent dynamics. It remains to be seen 
whether aspects of this picture could actually be astrophysically 
relevant.

Regarding the energetics, it should be noted that the mechanism 
described here does not lead to significant (ultra-relativistic) acceleration 
of the strings. While internal energy of the string can be converted to kinetic 
energy of motion along the axis of symmetry, the resulting $\Gamma$ factors 
in the examples we studied are rather close to unity, the largest velocity
we reported being $0.39 c$.  
Therefore, even if the mechanism we have found is somehow relevant to collimation, 
it  does not appear able to explain the acceleration
of relativistic jets. 
The acceleration would have to occur due to 
the presence of electromagnetic fields, or to some other mechanism. 
In particular, the process discussed here does not tap into the
rotational energy of the black hole, which is a prime candidate
for the source of ultra-relativistic jet energy. 

On the contrary, the collimation effect 
we find persists even at the Newtonian level, which may be 
advantageous: collimated jets seem to be present not only around 
black holes, but also around objects with significantly 
weaker gravitational fields, such as accreting young stars 
for instance~\cite{livio}. 
Therefore, if there is a common mechanism for the collimation, 
it should not depend on relativistic effects. Hence, 
the effects described here could perhaps be part of a more 
involved process that collimates and drives astrophysical jets.

\section{Conclusions}

We have studied the motion of relativistic, current carrying strings moving axisymmetrically
on the background of a Kerr black hole. First the equations of motion and
conserved quantities in the conformal gauge were found. 
Next,  we determined how the energy of a string 
that is instantaneously at rest depends on the current and the position of the
string. Contours of this quantity determine boundaries of the possible motion of a string with 
that energy and current. We compared this analysis with that for a Newtonian elastic ring 
moving around a point mass, and found that system to be qualitatively similar. 
 
By considering a number of examples and various plots of the relevant functions, 
the possible types of motion were mapped out. 
Regions of parameters for which the string falls into the black hole, or is trapped 
in a toroidal volume,  or can escape to infinity, were identified. After this general analysis,
we examined representative trajectories found by numerical integration, illustrating 
various interesting behaviors. In particular, we found that a string can start out at
rest near the equatorial plane and, after bouncing around, be ejected out along the axis,
some of its internal (elastic or rotational kinetic) energy having been transformed into 
translational kinetic energy. The resulting velocity can be an order unity fraction of the
speed of light.

Finally, we addressed the question of possible astrophysical significance of this
system as a simple model of MHD plasma flux tubes, which might conceivably
 play a role in jet formation and collimation.

\begin{acknowledgments}
We thank Vijay Kaul for collaboration in the early stages of this research.
This work was supported by the National Science Foundation under grant PHYS-0601800. 
\end{acknowledgments}


\end{document}